\documentclass[fleqn,usenatbib]{mnras}
\usepackage{newtxtext,newtxmath}
\usepackage[T1]{fontenc}

\DeclareRobustCommand{\VAN}[3]{#2}
\let\VANthebibliography\thebibliography
\def\thebibliography{\DeclareRobustCommand{\VAN}[3]{##3}\VANthebibliography}

\usepackage{graphicx}	
\usepackage{amsmath}	
\usepackage{orcidlink}  
\usepackage{soul}       

\newcommand{\Mcrit}{M_{200\mathrm{c}}}
\newcommand{\MSMBH}{M_{\mathrm{SMBH}}}
\newcommand{\Mstar}{M_{*}}
\newcommand{\Msun}{\mathrm{M}_{\odot}}
\newcommand{\betakin}{\beta}

\title[The star-forming past and quenching of BCGs]{The star-forming past and quenching of brightest cluster galaxies in TNG-Cluster}

\author[Sunna Gottschewski et al.]{Sunna Gottschewski$^{\orcidlink{0009-0008-5057-7138}1}$\thanks{E-mail: sunna.gottschewski@stud.uni-heidelberg.de},
Joey Braspenning$^{\orcidlink{0009-0003-3956-4890}1}$,
Annalisa Pillepich$^{\orcidlink{0000-0003-1065-9274}1}$,
Katrin Lehle$^{\orcidlink{0009-0005-5700-6483}2}$, and
Dylan Nelson$^{\orcidlink{0000-0001-8421-5890}2,3}$
\\
$^{1}$Max-Planck-Institut für Astronomie, Königstuhl 17, D-69117 Heidelberg, Germany 
\\
$^{2}$Universität Heidelberg, Zentrum für Astronomie, Institut für Theoretische Astrophysik, Albert-Ueberle-Str. 2, D-69120 Heidelberg, Germany
\\
$^3$Universit\"at Heidelberg, Interdisziplin\"ares Zentrum f\"ur Wissenschaftliches Rechnen, INF 205, 69120 Heidelberg, Germany
}

\date{Accepted XXX. Received YYY; in original form ZZZ}

\pubyear{\the\year{}}

\begin{document}
\label{firstpage}
\pagerange{\pageref{firstpage}--\pageref{lastpage}}
\maketitle

\begin{abstract}
The Brightest Cluster Galaxies (BCGs) in the most massive galaxy clusters of our Universe are, at present, mostly quiescent, elliptical, and red. However, their formation histories are unclear and likely different from those of massive ellipticals in lower-mass haloes. Simulations face the challenge of simultaneously requiring large volumes to sample massive clusters and sufficient resolution to model their stellar populations. Therefore, we make use of TNG-Cluster, one of the highest-resolution cosmological simulations of galaxy clusters, which returns 352 BCGs at $z=0$ with stellar masses $M_{*, <30\, \mathrm{pkpc}} = 10^{11.6-12.9} ~\Msun$ and total halo masses $\Mcrit = 10^{14.3-15.4}~\Msun$. 93 per cent of the BCGs are quenched at $z=0$, whereas at redshift $z>3$ all their main progenitors were still star forming. The star-formation histories of individual BCGs are complex, including multiple (re)quenching episodes. They rejuvenate on average $2.1 \pm 1.8$ times since $z=3$. BCGs quench for the first time $9.9^{+0.9}_{-1.4} ~\mathrm{Gyr}$ ago (10th-90th percentiles; $z=1.7^{+0.5}_{-0.5}$), while the last and final quenching event occurred $7.0^{+2.8}_{-5.0} ~\mathrm{Gyr}$ ago ($z=0.8^{+0.8}_{-0.6}$). We find a large scatter in the halo, stellar, and SMBH masses at both first and last quenching, suggesting no characteristic mass at which quenching occurs. Instead, BCGs quench for the first time when the cumulative kinetic energy injected by their SMBH equals the BCG binding energy, corresponding to a delay of $2.3^{+1.2}_{-1.2} ~ \mathrm{Gyr}$ after the first SMBH kinetic energy injection. Namely, TNG-Cluster BCGs quench when they experience significant gas expulsion. Our work shows that BCGs have complex and varied star-formation histories, with multiple rejuvenation episodes, but they ultimately quench due to SMBH feedback.

\end{abstract}



\section{Introduction} 
\label{sec:intro}

Brightest Cluster Galaxies (BCGs) are the giants of the galaxy population, ranking among the most massive and luminous galaxies in the Universe \citep{Matthews1964, Dubinski1998, Bai2014}. They reside near the centres of galaxy clusters, the largest virialized structures in the Universe \citep[total masses $\gtrsim10^{14-15} \, \Msun$,][]{Feretti2012}, each containing hundreds to thousands of galaxies \citep{Varela2009}. Since the first BCG was detected in the Coma cluster by \citet{Zwicky1951}, numerous other examples of exceptionally bright galaxies in galaxy clusters have been discovered \citep{Kluge2020}.

BCGs are located in the centres of galaxy clusters, which are not only crucial probes of cosmology and high-energy astrophysical processes \citep[e.g.][]{Corasaniti2021}, but also play a key role in the study of dark matter \citep{Hektor_2012}. Because they sit at the nodes of the cosmic web, the formation histories of BCGs are closely linked to the large-scale structure of the Universe \citep{chu2021}, and provide a valuable probe of environmental effects on galaxy evolution \citep{wetzel2014galaxy}.

At present day, BCGs generally host stellar populations exceeding $\gtrsim 10^{12} \, \Msun$ and have elliptical, bulge-dominated morphologies. They are mostly quenched or have low star-formation rates \citep[SFRs; e.g.][]{Aragon-Salamanca1998, Liu2012}, and consequently red stellar-light colours. 
Observations of optically selected BCG samples suggest that BCGs undergo rapid stellar mass growth between $z=1.5-0.5$ \citep{Lin2013}. However, using spectroscopic observations of BCGs in the SDSS footprint, \citet{Cerulo2019} find that only 9 per cent are star forming at $z < 0.1$. This fraction decreases rapidly towards lower redshift and higher cluster, i.e. halo mass. They find that massive BCGs ($\Mstar > 10^{11.5}~\mathrm{M_\odot}$) are quenched at all halo masses, whereas less massive BCGs rapidly change from having quenched fractions of only 20 per cent at cluster mass $ \approx 10^{14.0}~\mathrm{M_\odot}$ to 90 per cent at cluster mass $ > 10^{14.5}~\mathrm{M_\odot}$. Cross-matches with X-ray observations suggest that all star-forming BCGs reside in cool-core clusters, indicating a scenario where rapidly cooling gas contributes to star formation.
At the same time, BCGs are often observed to have large spatial offsets from the location of the X-ray peak, and large peculiar velocities, hinting at recent merger activity, which could spark star formation and drive stellar mass growth through accretion \citep{Coziol2009, Lauer2014, DePropris2021}.

Even though the majority of BCGs are quenched below $z=1-2$ \citep[e.g.][]{Aragon-Salamanca1998, Liu2012, McDonald2016}, a small fraction of BCGs show residual star formation at low redshift \citep{Liu2012, Belli2017}. 
Low redshift episodes of star formation in BCGs, commonly called rejuvenation, appear to be mostly short starbursts lasting less than $200 \, \mathrm{Myr}$ \citep{Pipino2009}, which in turn are thought to be triggered by a renewed supply of cold gas, coming, for instance, through inflows along cosmic filaments, internal mechanisms such as dynamical instabilities, and gas-rich mergers \citep[e.g.][]{webb2015, Belli2017, Pawlik2018}. Alternatively, as identified by \citet{Wang2010, Pipino2009, Liu2012}, so-called cooling flows may be the drivers of starbursts, as recent star formation in BCGs occurs mostly in cooling-flow clusters with cooling times shorter than $1 \, \mathrm{Gyr}$. 

In lieu of a good understanding of BCG star-formation histories in the literature, there is instead a larger body of work on massive elliptical field galaxies, specifically at Cosmic Noon ($z \sim 2-3$), when the cosmic star-formation rate peaked \citep[e.g.][]{Madau2014, Forster2020}. Observations of the most massive galaxies at Cosmic Noon show that the majority of them quenched for the first time at $z\approx2$ for $\Mstar \gtrsim 10^{10.2} \, \Msun$ \citep{Belli2019, Slob2024}, and $z\approx 3$ for $\Mstar \gtrsim 10^{11} \, \Msun$ \citep{Carnall2019, Slob2024}. In general, it is found that more massive galaxies quench earlier \citep{Forster2020}. In fact, for the most massive objects, the fraction of quiescent galaxies is known to grow rapidly between $z=1-2$ \citep[see e.g.][]{Carnall2019, Weaver2023, McDonald2016}.

These observed trends in quiescent fractions and star formation activity raise the question of which physical processes are responsible for suppressing star formation and ultimately quenching BCGs, the most massive galaxies in the Universe. While there are many possible pathways to quiescence in galaxies, \citet{Peng2010} has shown that massive galaxies are likely to quench due to internal factors, independent of their environment. 

Star formation is fundamentally tied to the availability of cold gas within a galaxy and has been observed to be regulated by outflows, both driven by star formation itself and by AGN feedback coming from the super-massive black hole (SMBH) residing in massive galaxies \citep[see reviews of][]{Somerville2015, Forster2020}. Observations agree that AGN-driven winds dominate in high-mass galaxies and, especially around Cosmic Noon, may be sufficiently strong to expel gas from galaxy cores, reducing the reservoir of cold, dense gas needed for star formation while simultaneously heating the gas \citep{Forster2020}.

Observational evidence supports this idea: many X-ray luminous AGNs have been observed to host ionized gas outflows capable of removing gas from the central regions of galaxies \citep[e.g.][]{webb2015, Rupke2017, Belli2019}. At low black hole accretion rates, SMBHs are believed to launch powerful jets from their accretion discs, which couple to the surrounding intracluster gas and eventually play a crucial role in the quenching, or at least in keeping galaxies quenched. In X-ray observations of massive galaxies, these jets have been observed to form large bubbles or cavities, showing suggestive evidence of their impact on the surrounding gas \citep{Birzan2004, Fabian2012}.

The critical role of AGNs and their feedback in galaxy formation and quenching has been investigated in several cosmological simulations of galaxies over the last decade. Simulations such as EAGLE \citep{Schaye2015, Crain2015}, Horizon-AGN \citep{Dubois2016}, IllustrisTNG \citep{Weinberger2017, Weinberger2018, Pillepich2018b, Nelson2018a}, SIMBA \citep{Scharre2024}, and ASTRID \citep{Ni2025} agree on the importance of AGN feedback in producing galaxy populations that agree with observations at the high-mass end ($\Mstar \gtrsim 10^{10-10.5} \, \Msun$). While they all find that AGN feedback is crucial for quenching star formation in high-mass galaxies, the details vary \citep[e.g.][]{Donnari2021obs}.

Modern high-resolution cosmological simulations, however, are typically either too small to have a significant population of massive clusters (such as EAGLE, TNG50/TNG100, and ASTRID) or have relatively low numerical resolution (such as FLAMINGO). That is why, in this work, we use the TNG-Cluster simulation \citep{Nelson2024} to quantify and study the star-forming past of the BCGs. 

TNG-Cluster is a spin-off project of the IllustrisTNG simulation suite \citep{Nelson2019b}\footnote{https://www.tng-project.org/} and consists of zoom-in magnetohydrodynamical cosmological galaxy simulations of galaxy clusters selected from a volume of  $\sim 1 \, \mathrm{Gpc}^3$, returning a large sample of 352 target haloes with masses $\Mcrit \gtrsim 10^{14.3-15.5} \, \Msun$ at $z=0$ \citep{Nelson2024}. All selected haloes are resimulated at the same resolution as TNG300, the largest volume of the IllustrisTNG suite, with gas cells and stellar particles of about $10^7\,\Msun$, providing a relatively high-resolution, large sample of massive clusters and their BCGs.

TNG-Cluster is based on the same code and galaxy-formation model as the IllustrisTNG simulations \citep{Weinberger2017, Pillepich2018b}. The IllustrisTNG model has been shown to produce galaxy populations that are consistent with a broad range of observational constraints across various physical processes, including star formation \citep{Donnari2019}, galaxy quenching \citep{Donnari2021obs}, and the colour bimodality of galaxies \citep{Nelson2018a}. Within such a model, the quenching of massive central and satellite galaxies occurs due to mechanical energy injections from SMBHs in kinetic and low-accretion mode \citep[e.g.][]{Weinberger2018, Nelson2018a, Zinger2020, Donnari2021, kurinchi-vendhan2024}, further indicating that AGN feedback is crucial for reproducing the level of galaxy quenching observations show \citep{Donnari2019, Terrazas2020,Donnari2021obs}.

Supported by the realism of the galaxies simulated with the IllustrisTNG model and by the enhanced statistics allowed by TNG-Cluster, in this work, we investigate the star-forming past of the BCGs in the TNG-Cluster simulation. Previous works based on the TNG300 simulation have shown that the star-formation rates of IllustrisTNG BCGs are in the ballpark of observational estimates \citep{Montenegro-Taborda2023}. In particular, the BCGs in TNG300 reach their maximum SFRs at $z \sim 3-4$ and decrease strongly at $z \sim 2-3$, becoming mostly quenched at $z \sim 1$ \citep{Montenegro-Taborda2023}. Building on these results, we extend the analysis to TNG-Cluster, whose sample of 352 BCGs spans a broader and more massive stellar-mass regime, $M_{*,<30,\mathrm{pkpc}} = 10^{11.6-12.9},\Msun$ at $z=0$. We investigate how many BCGs are quenched across cosmic epochs, how often BCGs rejuvenate during their evolution, and what drives their star-formation quenching. 

So, we first introduce the TNG-Cluster simulation and our methods in Section~\ref{sec:Methods}, followed by a detailed investigation of quenching events in BCGs in Section~\ref{sec:results}. This includes time evolution, gas distributions, and correlations between BCG and SMBH properties. Following this, in Section \ref{sec:BCGs_SMBH}, we study the physical mechanism that triggers quenching: SMBH feedback. We then discuss our results in the context of gas-rich mergers and cool core transformations in Section~\ref{sec:discussion}. Finally, we present our conclusions in Section~\ref{sec:conclusions}.


\section{Methods}
\label{sec:Methods}

\subsection{TNG-Cluster}
\label{sec:Methods_sims}

TNG-Cluster\footnote{http://www.tng-project.org/cluster/} \citep{Nelson2024} is a spin-off project of the original IllustrisTNG cosmological gravo-magnetohydrodynamical simulation suite \citep[TNG hereafter;][]{Pillepich2018a, Springel2018, Marinacci2018, naiman2018, Nelson2018a, Nelson2019_tng50, Pillepich2019_tng50}, which models a large range of physical processes relevant for galaxy formation by solving the coupled evolution of cold dark matter (DM), gas, stars, and SMBHs over time from shortly after the Big Bang ($z = 127$) to the present day \citep[$z=0$,][]{Nelson2019b}. 

TNG-Cluster consists of zoom-in simulations of 352 selected galaxy cluster-mass haloes from a $1~\mathrm{Gpc}^3$ simulation box, which are re-simulated with the IllustrisTNG model at TNG300 resolution ($m_{\mathrm{baryon}}=1.2 \times 10^{7} ~\mathrm{M_\odot}$). These high-resolution regions are embedded within the lower-resolution parent volume, preserving the large-scale cosmic environment. The zoom-in regions are selected such that, at $z=0$, the targeted galaxy clusters span halo masses of $M_{\rm 200c}$\footnote{$\Mcrit$ refers to the total mass enclosed within the radius $R_{200\mathrm{c}}$, within which the mean enclosed mass density is 200 times the critical density of the Universe.} $\approx 10^{14.3 - 15.4} \, \Msun$. Combined with the TNG300 simulation, TNG-Cluster provides a flat halo mass distribution up to $10^{15}$~M$_\odot$, after which the number of haloes is volume-limited by the parent simulation.

TNG-Cluster uses the IllustrisTNG galaxy-formation model \citep{Weinberger2017, Pillepich2018b} and employs the AREPO code \citep{Springel2010} with initial conditions based on the \citet{Planck2016} standard spatially-flat $\Lambda$CDM cosmology, characterized by the following 6 parameters: $\Omega_{\Lambda,0} = 0.6911$, $\Omega_{m,0} = 0.3089$, $\Omega_{b,0} = 0.0486$, $\sigma_8 = 0.8159$, $n_s = 0.9667$ and $h = 0.6774$.
Particles and cells are organized into groups using a standard friends-of-friends (FoF) algorithm \citep{Davis85} based on the DM distribution with linking length $b = 0.2$. These are then divided into gravitationally bound subgroups, or subhaloes, identified by the SUBFIND algorithm \citep{Springel2001}. Galaxies are subhaloes with non-vanishing galaxy stellar mass. In this paper, galaxy clusters are, broadly speaking, FoF-objects with total mass exceeding, e.g. about $\Mcrit=10^{14}$~M$_\odot$. 

In this work, we focus exclusively on the 352 zoom-in target haloes of TNG-Cluster and on cosmic times between $z=6.5$ and $z=0$. We identify the 352 BCGs of our sample as the first subhalo in each of the 352 haloes at $z=0$. Typically, a BCG is the most massive galaxy of a cluster and resides at the bottom of the cluster's potential well. We trace each BCG back in time using the main progenitor branch from the SUBLINK merger tree, which tracks the descendants of subhaloes by identifying common dark matter particles \citep{Rodriguez2015}. SUBLINK allows galaxies to skip snapshots when they are not identifiable. In these cases, we linearly interpolate the missing values.

As anticipated in Section~\ref{sec:intro}, it has been shown that in the IllustrisTNG model massive galaxies ($\gtrsim 10^{10.5}$ M$_\odot$ in stars) halt their star formation and become red thanks to the effects of SMBH feedback \citep{Weinberger2017, Nelson2018a, Donnari2021}, in particular in its kinetic low-accretion rate mode \citep{Zinger2020, Terrazas2020, Davies2019}, also at Cosmic Noon, i.e. at $z=1-4$ \citep{kurinchi-vendhan2024}. As this is also relevant to our study on BCGs, below we briefly summarize salient aspects of the underlying SMBH physics.

\subsection{SMBH physics and properties}
\label{sec:Methods_smbhs}

In TNG-Cluster, an SMBH with mass $1.7 \times 10^{6}~\mathrm{\Msun}$ is seeded into a halo whenever this halo exceeds a FoF mass of $7.38 \times 10^{10}~\mathrm{\Msun}$ and does not yet contain an SMBH \citep{Weinberger2017}. This SMBH then grows either through merging with other black holes during galaxy mergers or through the accretion of mass following the so-called Eddington-limited Bondi accretion rate \citep[see][]{Weinberger2017, Springel2005a}. For our analysis, for each BCG under consideration, we select the most massive SMBH within a sphere of 30 pkpc around the position of the BCG at each snapshot.

The accretion rate is used as a criterion for the SMBH feedback, which, in the TNG simulations, is implemented in two modes \citep{Weinberger2017, Weinberger2018, Nelson2019b}. At low-accretion rates, AGN feedback is implemented as kinetic energy injections, which are comparatively efficient at affecting the properties of the surrounding gas, by imparting momentum via discrete pulses in random directions \citep{Weinberger2017}. On the other hand, at higher accretion rates, SMBHs in the IllustrisTNG model switch to the so-called thermal mode SMBH feedback, which is less efficient and continuously injects thermal energy into the local environment \citep{Weinberger2018}. Previous work has shown that kinetic SMBH feedback results in a net momentum injection into the ICM around the galaxy and a change in the total gas energy, heating it and lowering cooling rates. It is therefore both actively quenching galaxies by ejecting gas \citep{Davies2019} and, at the same time, prevents subsequent star formation by suppressing cooling and inflows \citep{Zinger2020}.

The low- vs high-accretion rates that determine whether SMBHs do feedback in kinetic or thermal mode, respectively, are relative to the Eddington limit, as explained in \citet{Weinberger2017}. The Eddington limit is the theoretical accretion rate at which the pressure from the incoming gas equals the outflows in an equilibrium \citep{Weinberger2018}. 

A SMBH may be, at any given time, at low or high accretion rates depending on the conditions of the gas in its surroundings, which in turn depend on the properties and evolutionary stages of its host galaxy. Irrespective of the instantaneous mass accretion state and hence instantaneous feedback mode, for any SMBH, we can determine the cumulative energy ever injected in either feedback mode, e.g. for the kinetic one: 
\begin{equation}
    E_{\mathrm{kin, SMBH}} = \int_0^t \Dot{E}_{\mathrm{kin, SMBH}}(t') dt'.
\end{equation}
This sums up the energy of all kinetic energy injections up to a given point in time, and is a good proxy for the long-term, integrated effect of SMBH feedback. Additionally, we can determine the time at which the first kinetic energy injection from an SMBH occurred, which is when $E_{\mathrm{SMBH, kin}} > 0$ is true for the first time.

\subsection{Definitions}
\label{sec:Methods_defs}

In this section, we introduce common operational definitions used throughout this work.

\begin{figure*}
  \centering
    \includegraphics[width=0.85\linewidth]{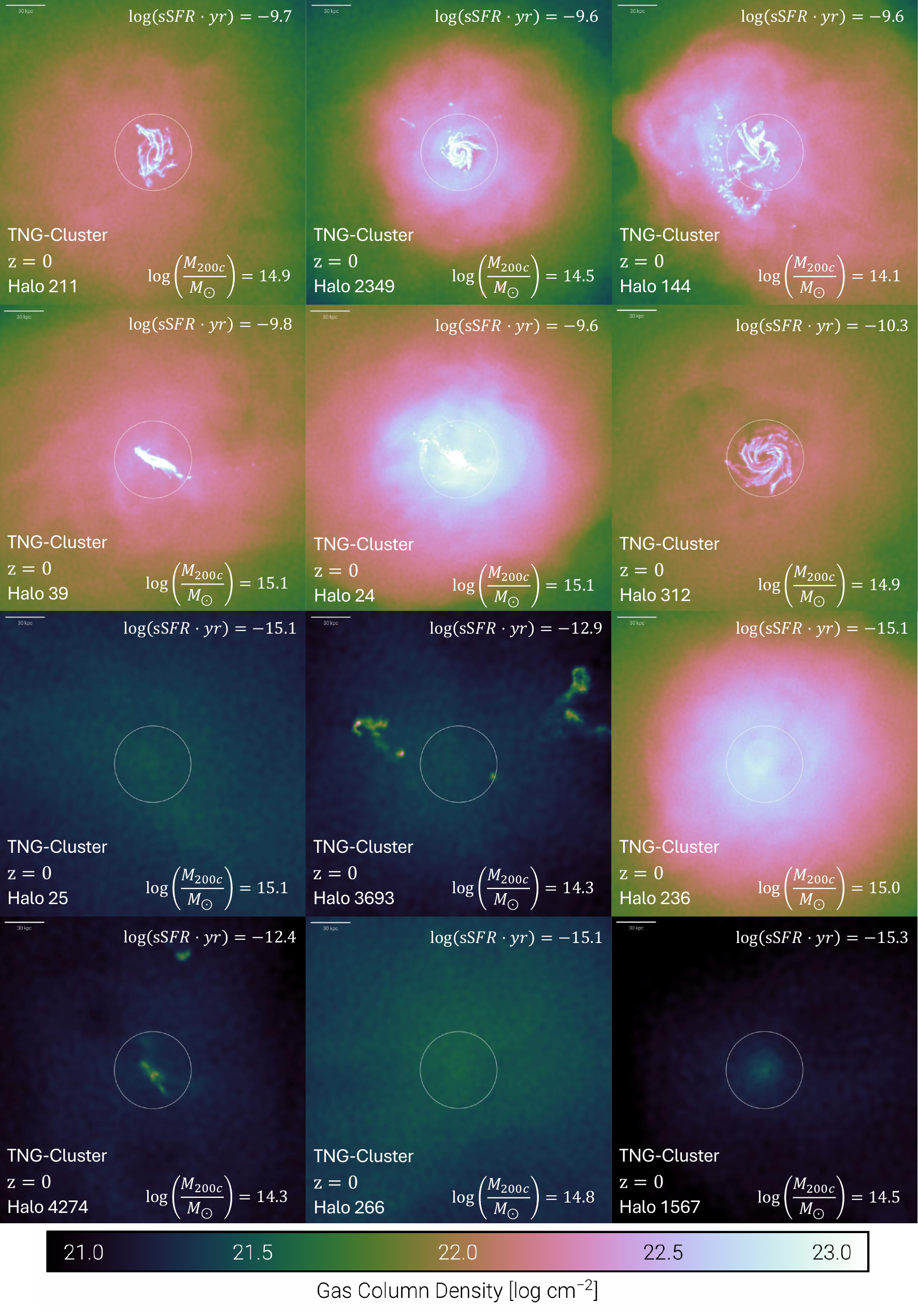}
  \caption{\textbf{Star-forming BCGs are discy.} Maps of the projected gas column density at $z=0$ for 12 example BCGs from TNG-Cluster. They are selected to cover the full range of halo masses and sSFR values. The upper two rows are BCGs that are star forming, while the bottom two rows depict quenched BCGs. The white circle represents the radius of 30 pkpc. sSFR and halo mass at $z=0$ are annotated. Star-forming BCGs show higher gas densities and, in most cases, characteristic gaseous disc features, while this is not the case for the quiescent galaxies. Maps generated with the TNG online visualization tool.}
  \label{fig:gasmaps}
\end{figure*}

\subsubsection{Properties of the BCGs and their host galaxy clusters}

We use $\Mcrit$ for the total halo mass of the host halo and $M_{*,<30 \, \text{pkpc}}$ for the stellar mass of the BCGs. The latter is within a 3D aperture of 30 physical kpc (hereafter pkpc) and is measured by summing the masses of all stellar particles from the snapshot data within that radius centered on the subhalo (galaxy) position.

We measure the instantaneous star-formation rate (SFR) of a BCG by summing over the individual SFRs of all gas cells within a 3D aperture of 30 pkpc around the subhalo position ($\mathrm{SFR}_{<30 \, \text{pkpc}}$) for every snapshot.
For the purpose of this work, all galaxies with SFR~$\equiv~0$, arising from the numerical resolution limit of the simulation, are assigned random values in the range of $\log(\mathrm{SFR}) = -3 \pm 0.5~\Msun/\mathrm{yr}$ dex.
The specific star-formation rate (sSFR) is then defined as the SFR per unit stellar mass of the galaxy.

We define the gas binding energy of a BCG following \citet{Terrazas2020} as
\begin{equation}
    E_{\text{bind, BCG}} = \frac{1}{2} \sum_{i(<30 \, \text{pkpc})}m_i \phi_i,
\end{equation}
where $m_i$ is the mass of the gas cell and $\phi_i$ is the gravitational potential at the cell's position.

We can measure the quantities above at a time cadence that is much coarser than the time resolution of the actual simulations: namely, at 14 points in time between $z=6.5$ and $z=0$.
In the case of 36 BCGs (corresponding to 10 per cent of the BCG sample), there are problems with the gas potential values in the output -- these BCGs are excluded from all analyses, including the binding energy. This does not affect the findings of this work.

\subsubsection{Quenched definitions and rejuvenation times}

There is no universal definition for categorizing galaxies into star-forming and quenched. A common method is to set an sSFR threshold, labelling galaxies with sSFR above this value as star-forming, independent of redshift. However, this criterion does not take into account the time evolution of the star-forming main sequence (SFMS) in the SFR-$M_*$ plane \citep{Whitaker2012} and is therefore only applicable over a limited redshift range \citep{Donnari2019}.

Taking this effect into account, in this work, galaxies are classified by comparing their sSFR to the Hubble time $t_H(z)$ at the redshift of interest, as in \citet{Tacchella2019}. The Hubble time approximates the age of the Universe under the assumption of a constant expansion rate. Galaxies with an sSFR lower than
\begin{gather}
    \mathrm{sSFR}_{\text{quiescent}}(z) = \frac{1}{6 \cdot t_H(z)}
    \label{eq:sSFR}
\end{gather}
are identified as quenched. This criterion means that, for quenched galaxies, it would take more than 6 times the age of the Universe to double their stellar mass, which means that they effectively do not grow through star formation.

The most relevant values for investigating the star-forming past of BCGs are the times at which quenching and rejuvenation occurred and the durations of quenching and rejuvenation episodes. A quenching (rejuvenation) event happens each time the $\mathrm{sSFR}_{<30 \, \text{pkpc}}$ of a BCG crosses the quenching-limit $\mathrm{sSFR}_{\text{quiescent}}(z)$ (Eq.~\ref{eq:sSFR}) from above (below). Because this can happen between two snapshots, we determine the time of quenching using linear interpolation, weighting the two snapshot times by their respective sSFR values. We impose a minimum threshold of 5 Myr for the duration of quenching and rejuvenation episodes, in order to minimize dependencies on the interpolation scheme between consecutive data snapshots. This choice is also physically motivated by the response timescales of star formation indicators, such as H$\alpha$ emission, which are sensitive to changes in star formation activity over timescales of about $5-10$ Myr.

The time of first quenching is referred to as the age of the Universe when the galaxy undergoes its first quenching in its evolution, while at the last quenching event, the galaxy quenches for good and remains quenched until $z=0$. BCGs that are star-forming at $z = 0$ are not considered in the analysis of the last quenching time.


\section{Results, according to TNG-Cluster}
\label{sec:results}

We hereby present the results from our study of the star-forming past of BCGs according to the TNG-Cluster simulation. In Section~\ref{sec:BCG_ssfr_present} we show that BCGs are mostly quenched at $z=0$, followed by their diverse star-forming past in Section~\ref{sec:BCG_ssfr_past}. We then explore the cause of quenching in BCGs in Section~\ref{sec:BCG_quenching_time}, where we correlate the time of quenching with halo and galaxy properties. In a subsequent Section~\ref{sec:BCGs_SMBH}, we discuss the physical mechanism that halts star formation in BCGs.

\subsection{BCGs are mostly quenched at present time} \label{sec:BCG_ssfr_present}

TNG-Cluster naturally returns a diverse population of BCGs at $z=0$. 
In Fig.~\ref{fig:gasmaps}, we show gas column density maps of a subset of 12 BCGs at $z=0$. The first two rows show star-forming BCGs, the last two rows show quenched BCGs. The white circle indicates the radius ($30~\mathrm{pkpc}$) used for our measurements (Section~\ref{sec:Methods_defs}), illustrating that this radius covers the inner part of the BCGs where dense gas is present, and star formation can occur. The sSFR values and halo masses at $z=0$ are annotated at the bottom of each panel, and the BCGs are selected such that they span the whole mass range of our BCG sample.

It can be easily seen that star-forming BCGs have high gas column densities, typically between $10^{22.5}-10^{23} ~ \mathrm{cm}^{-2}$ with face-on and edge-on disc characteristics. For the quenched galaxies, the densities typically do not exceed $10^{22}~ \mathrm{cm}^{-2}$, and there are no disc features. This aligns well with the fact that star formation is fueled by gas, and that haloes in TNG-Cluster with ongoing star formation have abundant reservoirs of cool gas \citep[see also][]{Rohr2025, Staffehl2025}. When this gas is missing, as is the case in the two bottom rows, the sSFR values are low, and the galaxies are considered quenched.
An exception to this can be seen in the third row, where halo 236 shows exceptionally high gas densities despite being quenched. We will see in the following section that this BCG quenched very recently.

\begin{figure*}
    \centering
    \includegraphics[width=\linewidth]{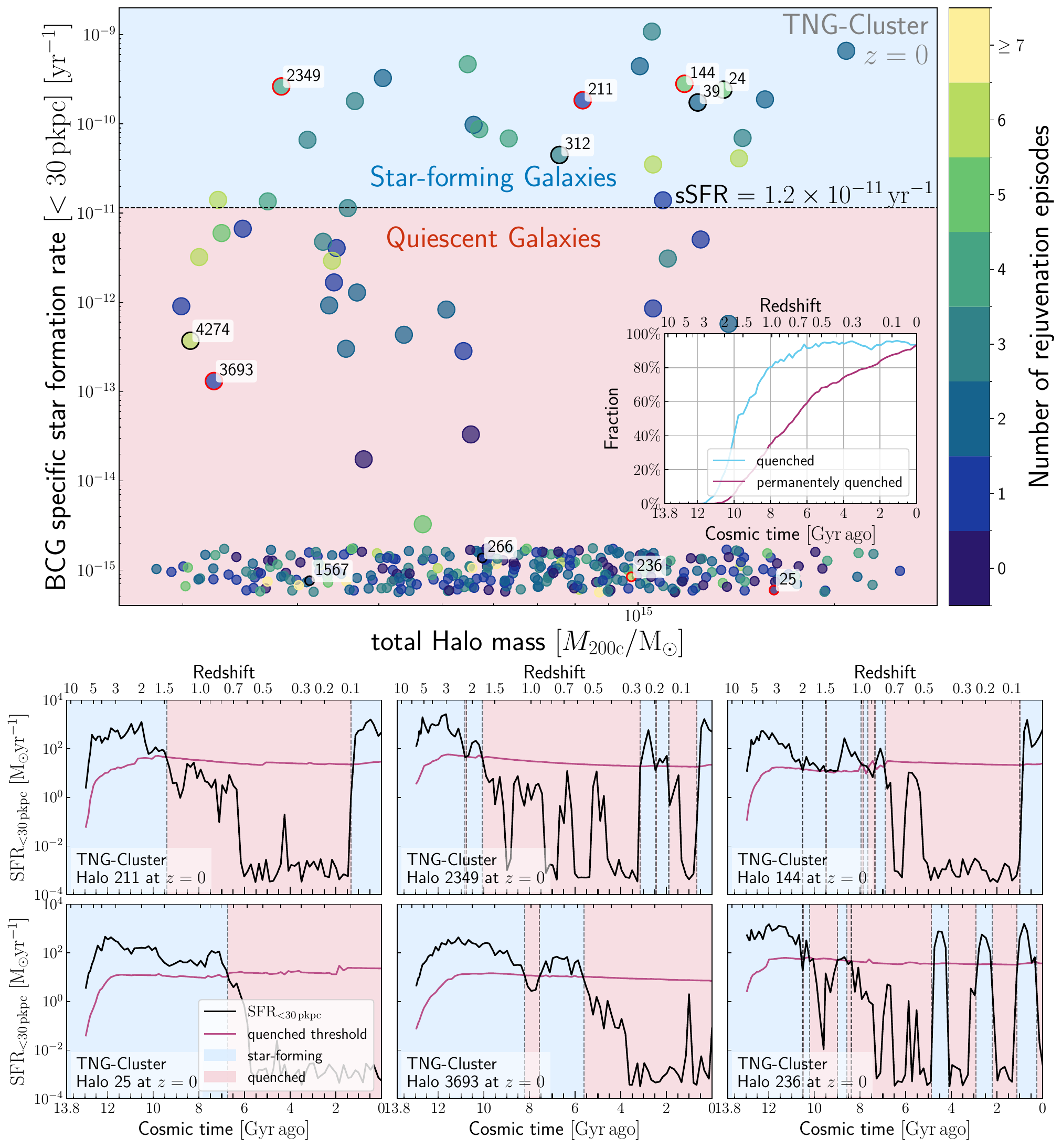}
    \caption{\textbf{According to TNG-Cluster, BCGs undergo multiple quenching and rejuvenation episodes.} Top: sSFR versus total cluster mass $\Mcrit$ of all BCGs in TNG-Cluster at $z=0$, colour-coded by the number of rejuvenation episodes any given BCG has undergone during its lifetime. The dashed line marks the $z=0$ quenched threshold $\mathrm{sSFR} = 1.21 \times 10^{11} \, \text{yr}^{-1}$, dividing the star-forming from the quiescent population. While 93 per cent of the BCGs are quenched, the number of rejuvenation episodes does not correlate with sSFR or halo mass. The BCGs shown in the gas maps in Fig.~\ref{fig:gasmaps} are circled in black and red, and their ID is annotated. In the inline plot, we plot the fraction of quenched galaxies over time. The blue curve corresponds to the total fraction of quiescent BCGs at each snapshot, the magenta curve represents the fraction of BCGs that are permanently quenched, meaning that they have already undergone their last quenching event and remain quenched until $z=0$. Both curves start growing from zero between 11 and 11.7 Gyr ago, i.e., between $z \sim 3$ and $\sim 2.3$. Bottom: Star-formation histories of 6 example BCGs, marked by the red circles in the scatter plot above, and selected such that they represent the whole variety of number of rejuvenation episodes while spanning the whole mass range. The 3 upper BCGs are star forming, whereas the 3 bottom BCGs are quenched at $z=0$. The black curve denotes the instantaneous SFR at each snapshot, while the magenta curve represents the limit below which the BCG is categorized as quenched. The blue (red) shaded areas denote when the BCG is star-forming (quenched). All BCGs begin as star-forming galaxies, while the frequency and duration of the rejuvenation periods differ considerably among BCGs.}
    \label{fig:sfr}
\end{figure*}

The top panel of Fig.~\ref{fig:sfr} shows the sSFR at $z=0$ of all BCGs as a function of present-day halo mass $\Mcrit$, i.e. total cluster mass. The grey dashed line represents the $z=0$ quenched limit $\mathrm{sSFR} = ~1.21 \times 10^{11}~ \mathrm{yr}^{-1}$, as per Eq.~\ref{eq:sSFR}. BCGs with sSFR below this value are quenched, while sSFR values above mean that they are star forming. The band of smaller scatter points at the bottom of the figure represent BCGs with an sSFR value below the resolution limit and are therefore artificially scattered around $\mathrm{sSFR} = 10^{-15} \, \text{yr}^{-1}$ -- they are all quenched.

Confirming results by \cite{Nelson2024}, we find that 93 per cent of the BCGs are quenched at $z=0$, corresponding to 329 of the 352 total BCGs in the sample, and aligning well with expectations and observations of a high fraction of BCGs being dead and red at recent cosmic epochs.

The values of sSFR clearly mirror the gas density maps shown in Fig.~\ref{fig:gasmaps}, with the 12 BCGs from that figure circled in red and black in the top panel of Fig.~\ref{fig:sfr}.

\subsection{BCGs have not always been quenched} \label{sec:BCG_ssfr_past}

Even though most BCGs are currently quenched, both in the Universe and in TNG-Cluster, this was not always the case in the past. In Fig.~\ref{fig:sfr}, the colour-coding represents how many rejuvenation episodes the BCGs have undergone during their evolution. On average, the BCGs rejuvenate 2.1 times, with 66 per cent undergoing between 1 and 3 rejuvenation episodes. 17 per cent of the BCGs quenched only once, for good and never rejuvenate, while a small fraction of 2 per cent of the BCGs have very turbulent lives with between 7 and 10 rejuvenation episodes. From Fig.~\ref{fig:sfr}, no strong correlation seems to be in place between the number of rejuvenation episodes and the cluster mass or sSFR at $z=0$.

\subsubsection{The fraction of quenched BCGs grows with time}
\label{subsubsec:quenched_fraction_grows}
The inset plot of Fig.~\ref{fig:sfr} quantifies the time-evolution of the fraction of quenched BCGs, as a population. The blue curve is the fraction of quenched galaxies at each time, while the magenta curve represents the fraction of permanently-quenched BCGs, which means that they do not have any later rejuvenations.

We find that at $z>3$, corresponding to a cosmic time of $11.7~ \mathrm{Gyr}$ ago, all BCGs of TNG-Cluster were star forming. After that time, the fraction of quenched galaxies starts to increase. While the fraction of permanently-quenched BCGs increases monotonically and almost linearly with time, the instantaneous fraction of quenched galaxies grows quickly between $z = 3$ and $z=1$ up to a fraction of $\sim90$ per cent and then stays on a plateau close to the value at present day, which is 93 per cent.

In the fraction of permanently-quenched BCGs, we see a change in slope at $z = 0.5$ (corresponding to $t \sim 5.3~\mathrm{Gyr}$ ago), after which the fraction starts to grow more slowly compared to before.
The gap between the blue and magenta curves can be interpreted as the fraction of BCGs that have quenched at least one time, but are currently star forming. The gap grows until $z\sim1$ and then narrows over time, implying that there are fewer star-forming, i.e., rejuvenation episodes with cosmic time.

\subsubsection{BCGs may undergo many rejuvenation episodes}

In the bottom panels of Fig.~\ref{fig:sfr}, we show the star-formation histories of a selected sample of 6 BCGs, the top row being star forming and the bottom row being quenched at $z=0$. We also clearly indicate the time evolution of the SFR (black) and the quenched threshold (magenta, see Eq. \ref{eq:sSFR}) of individual BCGs over cosmic time. The blue and red-shaded areas indicate periods when the BCG was star forming or quenched, respectively. The TNG halo ID\footnote{The GroupOrigHaloID from the TNG-Cluster halo catalogues.} is annotated in the plots, and the 6 selected BCGs are marked in the top panel by red circles.

We see that the star-formation histories naturally differ in their individual characteristics. They range from relatively calm evolutions, as in halo 211 (top left), with long star-forming and quenched periods, to highly episodic star-formation histories, such as halo 144 and halo 236 (right column), which undergo many rejuvenation episodes of varying durations. 

If we look back at the gas map at $z=0$ (Fig.~\ref{fig:gasmaps}) of halo 236, which just recently quenched $0.3~\mathrm{Gyr}$ ago, we see that it still contains a high amount of gas in its core despite being quenched. This suggests that the quenching of star formation does not always have to happen exactly simultaneously with the complete removal of the gas that fuels it, and can even precede it.

\subsubsection{Rejuvenation episodes become less frequent at lower redshift}

\begin{figure}
    \centering
    \includegraphics[width=\linewidth]{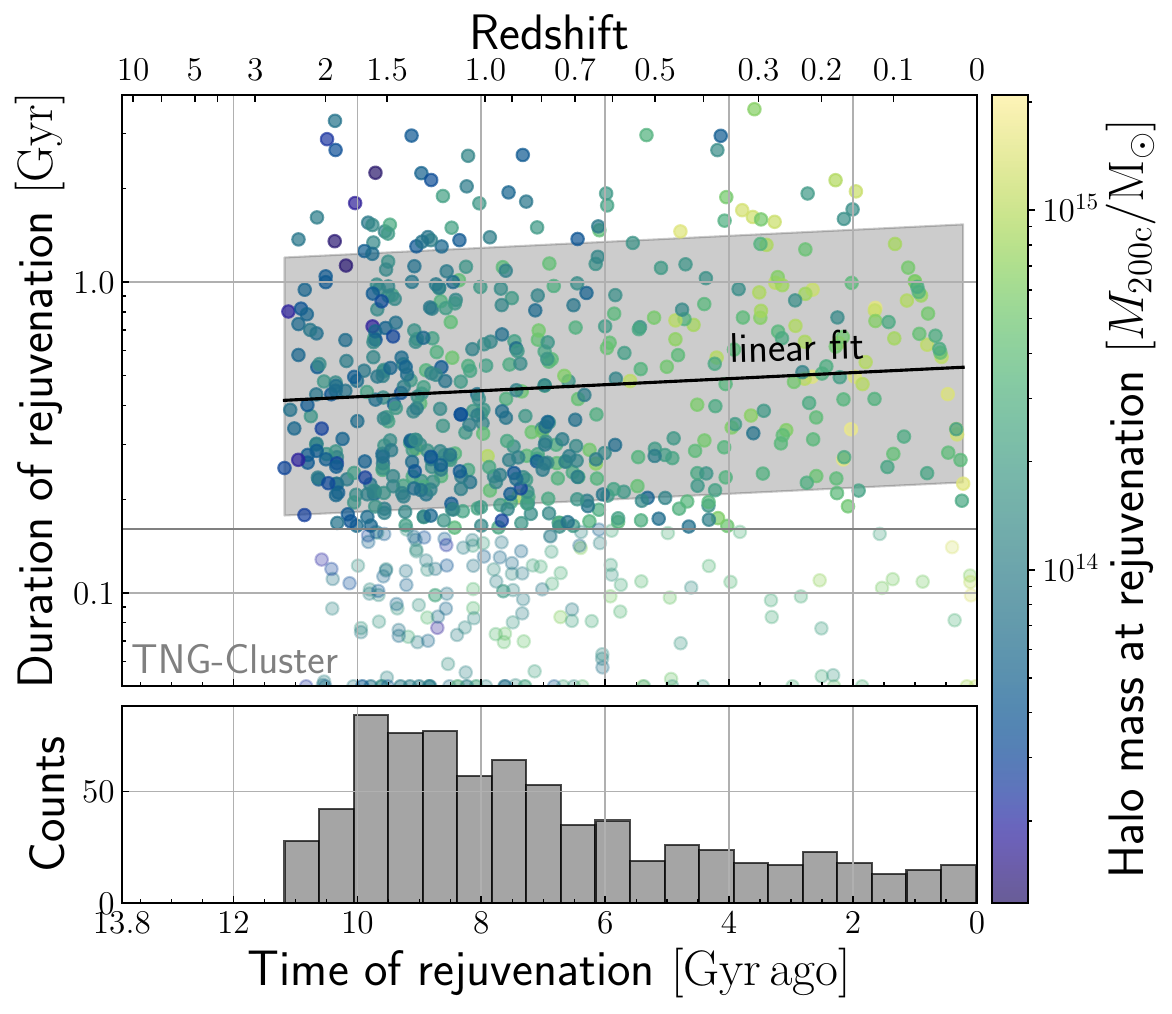}
    \caption{\textbf{According to TNG-Cluster, rejuvenation events in BCGs become less frequent with cosmic time.} Top: The time of all rejuvenations the BCGs undergo versus the duration of each rejuvenation episode, colour-coded by the halo mass $\Mcrit$ at the time of rejuvenation. Rejuvenations last on average for $500 ~\mathrm{Myr}$. The black curve represents the linear fit to the logarithmic duration with the 10th to 90th percentile of the scatter. Rejuvenation episodes lasting shorter than $160 ~\mathrm{Gyr}$, which corresponds to the approximate cadence of the snapshots, are excluded from the linear fit and the histogram below. The duration of rejuvenation episodes stays roughly constant over cosmic time, with a weak trend for longer episodes at more recent times. Even BCGs hosted by very massive halos with $\Mcrit>10^{15} ~\mathrm{\Msun}$ rejuvenate. Bottom: Histogram of the number of rejuvenation events over cosmic time. At $z \approx 1.5$, the number of rejuvenations is highest and then falls off. It stays roughly constant at a low level after $z\approx 0.5$.}
    \label{fig:rejuvenation_vs_duration}
\end{figure}

In the top panel of Fig.~\ref{fig:rejuvenation_vs_duration}, we show the duration of rejuvenation events as a function of the cosmic time at which they started. The colour represents the halo mass of the host haloes $\Mcrit$ at the time of rejuvenation. Rejuvenation episodes shorter than $160~\mathrm{Gyr}$, i.e. points below the grey horizontal line, are excluded from our analysis. Those episodes are shorter than the typical snapshot cadence, and therefore their precise duration may be influenced by the adopted interpolation scheme. We fit a linear relation to the logarithm of the duration, with shaded regions indicating the 10th to 90th percentiles around the linear fit.

We see that, according to TNG-Cluster, the duration of rejuvenation episodes stays approximately constant over cosmic time, with a weak trend towards longer episodes at more recent times. The rejuvenation events last on average $\approx 420 ~\mathrm{Myr}$ at $z = 2$ and $ \approx 530~\mathrm{Myr}$ at $z=0$. However, there is a large scatter with a few rejuvenation events lasting longer than $3 ~\mathrm{Gyr}$, and a large number lasting $\approx200~\mathrm{Myr}$.

We see that even BCGs in very massive clusters with $\Mcrit>10^{15}~\mathrm{\Msun}$, indicated by the yellow coloured points, can rejuvenate below $z = 0.5$, i.e. in the last 5 billion years.

Even though we don't find a strong trend in the duration of the rejuvenation episodes, the histogram in the lower panel of Fig.~\ref{fig:rejuvenation_vs_duration} demonstrates that rejuvenation events become less frequent with cosmic time. Most rejuvenations happen between $z = 1.5-1.0$, after which the number decreases and settles at a roughly constant rate between $z = 0.5 - 0.0$. This is in line with the quenched fraction of BCGs increasing with time and with the interpretation we gave to the evolution of the fraction of quenched BCGs of the inset plot of Fig.~\ref{fig:sfr} (see Section~\ref{subsubsec:quenched_fraction_grows}).

\subsection{BCGs quenched for good about 7 billion years ago}
\label{sec:BCG_quenching_time}

Amid such complex star-formation histories and a predominantly-quenched $z=0$ population, when did BCGs quench for the first time and when did they become permanently quenched? Does TNG-Cluster predict any obvious correlation between such times and halo, galaxy and SMBH properties?

\begin{figure*}
    \centering
    \includegraphics[width=1\linewidth]{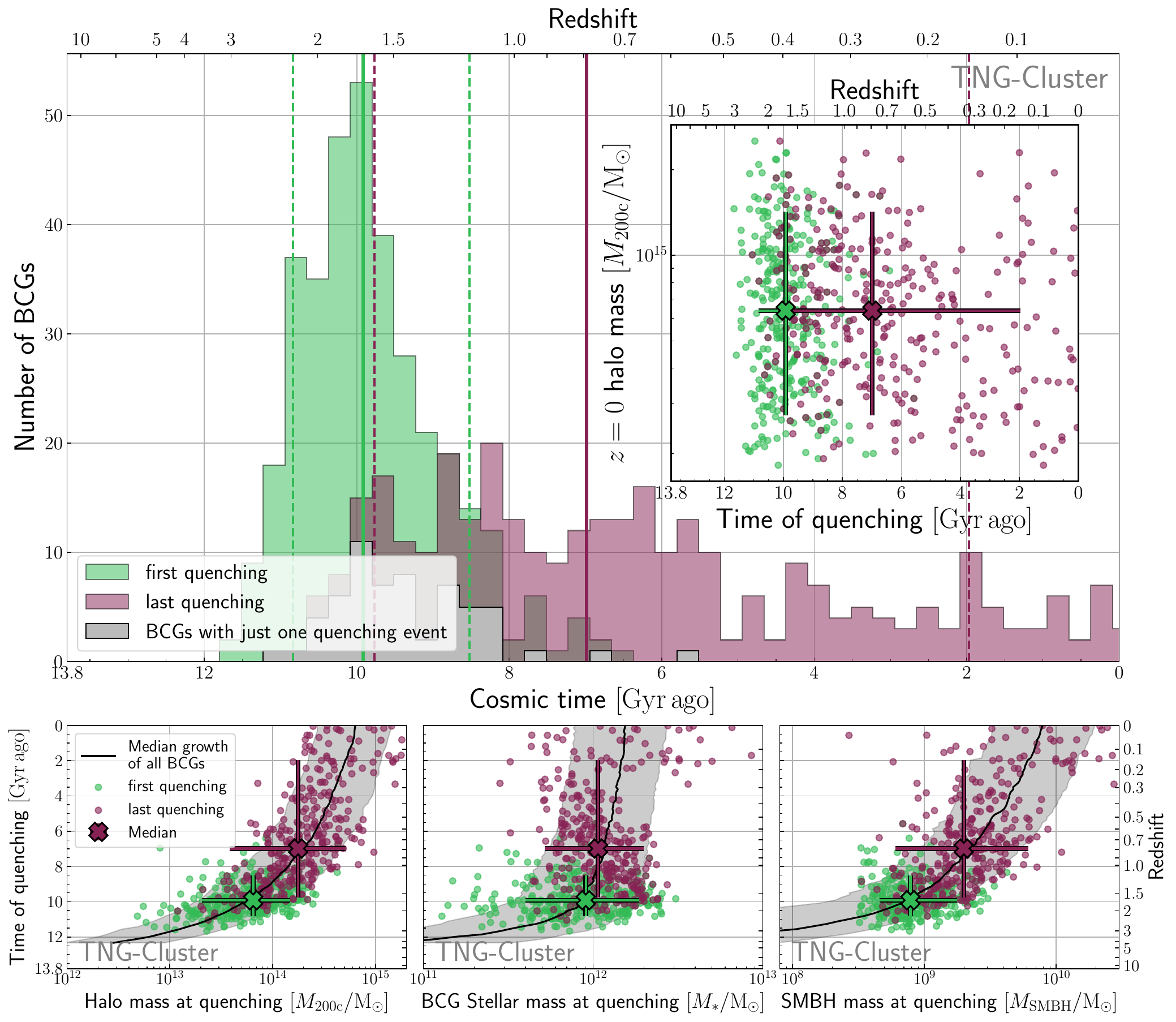}
    \caption{\textbf{When did BCGs quench according to TNG-Cluster?} Top: Distribution of the times of first and last quenching. The green (magenta) bars indicate the times of first (last) quenching of all BCGs; the grey bars represent BCGs with just one quenching, meaning that the first and last quenching events coincide. The solid vertical lines are the median, with the dashed lines denoting the 10th and 90th percentiles. The first quenching happens at a cosmic time of $9.9^{+0.9}_{-1.4} \, \mathrm{Gyr}$ ago, and the last quenching at $7.0^{+2.8}_{-5.0} \, \mathrm{Gyr}$ ago, meaning that the last quenching is more scattered than the first quenching. In the inline plot, the time of the first (last) quenching is shown in green (magenta) for all BCGs as a function of halo mass at $z=0$. The $z=0$ halo mass does not seem to correlate with the times of the first and last quenching. Bottom: Time of first and last quenching versus the halo mass $\Mcrit$ (left), stellar mass $\Mstar$, and mass of the SMBH $\MSMBH$, all at the corresponding quenching times. The median and the 10th-90th percentiles are plotted. The black curve shows the median mass growth across all halos, with the shaded region being the 10th to 90th percentile. The scatter in halo, stellar, and SMBH masses at the times of quenching is large and follows roughly the median mass growth. This suggests that, at least for BCGs, there is no characteristic mass of quenching.}
    \label{fig:quenchingtimes}
\end{figure*}

\subsubsection{The quenching times of BCGs}

In the top panel of Fig.~\ref{fig:quenchingtimes}, we present the quenching times of the BCGs in TNG-Cluster. The histogram shows, in green, the time of the first quenching of the BCGs, and in magenta, the time of the last quenching. We denote with grey the BCGs that undergo just one quenching event, meaning that the first and last quenching events are the same. The solid vertical lines indicate the median values, while the dashed lines mark the 10th and 90th percentiles of the quenching times distributions (which we quote hereafter).

According to TNG-Cluster, BCGs underwent their first quenching event about $9.9^{+0.9}_{-1.4} \, \mathrm{Gyr}$ ago (10th to 90th percentile), corresponding to $z=1.7$, on average. We see that the first quenching is quite synchronized between all BCGs and happens within a narrow time span of only $2.3~\mathrm{Gyr}$, or, in other words, 80 per cent of all BCGs quench for the first time between $z=2.2$ and $z=1.2$.

This contrasts with the last quenching event, which occurs $7.0^{+2.8}_{-5.0} \, \mathrm{Gyr}$ ago, corresponding to $z= 0.8$ on average, with 80 per cent of the BCGs quenching between $z=1.6$ and $z=0.2$. The last quenching events show a distribution with two plateaus: a higher plateau between $z=1.5$ and $z=0.5$, after which a second, lower plateau stretches all the way to the present day. The transition at redshift $z=0.5$ between these two plateaus aligns well with the turning point in the fraction of permanently-quenched BCGs in the inline plot in Fig.~\ref{fig:sfr}. 

The BCGs that undergo no rejuvenation episodes at all correspond to 17 per cent of the total population. For those BCGs, the quenching happens on average at $9.4^{+0.8}_{-1.2} \, \mathrm{Gyr}$ ago, which is a bit later than the average time considering all BCGs. This shows that BCGs that never rejuvenate have their first, and only, quenching event somewhat later than those in the rest of the population.

\subsubsection{Halo, galaxy, and SMBH masses at the times of quenching}

In the inset plot in Fig.~\ref{fig:quenchingtimes}, the times of the first and last quenching of all BCGs are plotted in green and magenta, respectively, versus the halo mass these BCGs have at $z=0$ on the y-axis. The median, 10th, and 90th percentiles (measured separately in the two 1-dimensional distributions) are shown by the crosses and error bars. The wide scatter in time of the last quenching compared to the first quenching event is clear from the much larger extent of the horizontal magenta error bar. By construction and selection, our population of BCGs have masses of $\log(M_{\rm 200c}/\Msun) = 14.8 ^{+0.3}_{-0.4}$ at $z=0$. Given the large spread in the data points, it is clear that the times of both first and last quenching do not correlate with $z=0$ halo mass. However, a weak trend may be in place, with a preference for BCGs residing in somewhat more massive haloes for more recent last quenching times.

In the three bottom panels of Fig.~\ref{fig:quenchingtimes}, we visualize the quenching time versus the mass of different components at the time of quenching: total cluster mass, i.e. halo mass $\Mcrit$, BCG stellar mass $\Mstar$, and SMBH mass $\MSMBH$. The median mass growth for all BCGs is shown by the black curve, with the shaded region denoting the 10th to 90th percentile.

In the leftmost bottom panel, we find that the first quenching occurs on average at halo masses of $\log(\Mcrit/\Msun) = 13.8 ^{+0.4}_{-0.5}$. The last quenching happens at halo masses of $\log(\Mcrit/\Msun) = 14.2 ^{+0.5}_{-0.7}$, which means that there is a significant overlap between the halo mass of the most massive haloes at first quenching and the least massive ones at last quenching. The halo masses at quenching largely follow the median mass evolution, including the degree of scatter, indicating that, in terms of halo mass, quenching halos are not special compared to the whole population. We further note that the halo mass at first quenching shows only a very shallow dependence on time of quenching (given the tight distribution of those quenching times across the BCG sample), as opposed to the mass at last quenching, which strongly increases with time of quenching, being in line with the median mass growth. The haloes that quench closest to $z=0$ have the largest halo masses at quenching, as noted in the inset panel above. This indicates that even very massive haloes can undergo rejuvenation events at low redshifts. Interestingly, between the first and the last quenching event, haloes grow on average 0.5 dex in mass.

In the middle bottom panel of Fig.~\ref{fig:quenchingtimes}, the stellar masses at first and last quenching are shown. According to TNG-Cluster, BCGs quench for the first time at $\log(M_{*}/\Msun) = 11.9 ^{+0.3}_{-0.4}$ and for the last time at $\log(M_{*}/\Msun) = 12.0 ^{+0.3}_{-0.3}$. Both quenching events show a wide scatter of 0.7 and 0.6 dex in $\Msun$, respectively, indicating that also for $\Mstar$ there is no characteristic mass of quenching, except that all BCGs are quenching at $>10^{11}~\mathrm{M_\odot}$. The stellar masses at quenching largely follow the median mass evolution. Even more so than for the total cluster mass, the stellar masses at first quenching seem to be independent of quenching time, while at the last quenching, there is a weak anti-correlation at $z\gtrsim0.3$: BCGs that quench later in time have a lower stellar mass $\Mstar$ and are clearly less massive than the median BCG between $z \sim 0.6-0.3$. However, we find a sudden transition in which, at $z\lesssim0.3$, the stellar masses are up to one order of magnitude higher and widely scattered. It is unclear what causes this effect, but it seems associated with the special subset of very massive haloes that quench for the last time at very recent times (data points in the top-right region of the leftmost bottom panel).

Whereas the smallest BCG mass at first quenching is just above $10^{11}~\mathrm{M_\odot}$, by the time of last quenching, there are no BCGs with stellar masses below $10^{11.5}~\mathrm{M_\odot}$. The interpretation of these trends is complicated by the fact that massive galaxies assemble their stars not only by converting their gas into stars but also by accreting stars via mergers and stripping, this being the case also for TNG-Cluster BCGs \citep[e.g.][for TNG]{Pillepich2018a, Montenegro-Taborda2023} -- we will get back to this in the Discussion.

Finally, in the bottom-right panel of Fig.~\ref{fig:quenchingtimes} we show the distribution of SMBH masses at the time of quenching, which again roughly follows the median trend. Typically, the first quenching happens when the central SMBHs have masses of $\log(M_{\mathrm{SMBH}}/\Msun) = 8.9 ^{+0.4}_{-0.2}$, i.e. with a scatter over $0.5~\mathrm{dex}$. The SMBH masses at the last quenching are even more diverse: $\log(M_{\mathrm{SMBH}}/\Msun) = 9.3 ^{+0.5}_{-0.5}$. We see a similar effect as for stellar mass, where the late quenching BCGs ($z \lesssim 0.3$) show a markedly increased scatter of their SMBH masses.

Given the distributions of Fig.~\ref{fig:quenchingtimes}, we do not identify a characteristic $\MSMBH$-threshold at which quenching occurs, nor in halo or galaxy mass. This result is apparently in contrast with the findings of e.g. \cite{Terrazas2020, kurinchi-vendhan2024}, who measure a sharp drop in the sSFR values at a certain SMBH mass-threshold for TNG100 and TNG300 galaxies, both at low and high redshifts, but for lower-mass systems.

\section{Kinetic SMBH feedback as the quenching mechanism also of BCGs}
\label{sec:BCGs_SMBH}

The IllustrisTNG AGN feedback in kinetic mode has been shown to be responsible for (triggering the) quenching of massive galaxies ($\gtrsim 10^{10.5} ~\Msun$ in stars) by removing gas via the generation of outflows  \citep{Nelson2018a, Weinberger2018, Terrazas2020, Zinger2020, Davies2020, Donnari2021}. More specifically, quiescence in TNG massive galaxies sets in once the cumulative energy injected by the low-accretion-rate kinetic AGN feedback exceeds the gravitational binding energy of the gas within the galaxy \citep{Terrazas2020}, naturally leading to inside-out quenching \citep{Nelson2019_tng50, EricaNelson2021}. 
At the same time, the kinetic mode of AGN feedback has been found to have a preventative effect, restraining gas inflows and cooling from reaching the ISM, and thereby impairing star formation \citep[e.g.][]{Weinberger2017, Zinger2020, Rohr2025}. In contrast, in the IllustrisTNG framework, thermal-mode SMBH feedback, operating at high accretion rates, has been shown to be significantly less effective at driving outflows and regulating the thermodynamical properties of the ISM and circumgalactic gas: it is, in fact unable to produce, alone, a red, quenched population of massive galaxies consistent with observations \citep{Vogelsberger2014, Weinberger2017, Nelson2018a, Pillepich2021}. 

Consistent with the larger gas reservoirs available at high redshift, seeded black holes in TNG initially grow rapidly through high Eddington-ratio accretion while operating in the thermal SMBH feedback mode. As the SMBH mass increases, the feedback transitions to the low-accretion kinetic mode \citep[e.g.,][]{Weinberger2017, truong2021}, which is substantially more effective at suppressing gas cooling, regulating the gas supply, and quenching star formation, while maintaining the SMBH in a self-regulated, low-accretion state.

This switch typically happens around $M_{\mathrm{SMBH}}~\approx 10^{8} ~\mathrm{M_\odot}$ at $z\sim0$, \citep{Weinberger2017, Weinberger2018}. However, an SMBH may in principle transition between accreting at low and high accretion, and hence between kinetic and thermal mode feedback,   depending on the physical local conditions and global properties of its host galaxies. 
At higher masses, once the transition has occurred, SMBHs typically remain predominantly in kinetic feedback mode and infrequently switch back to thermal mode.

\begin{figure}
    \centering
    \includegraphics[width=\linewidth]{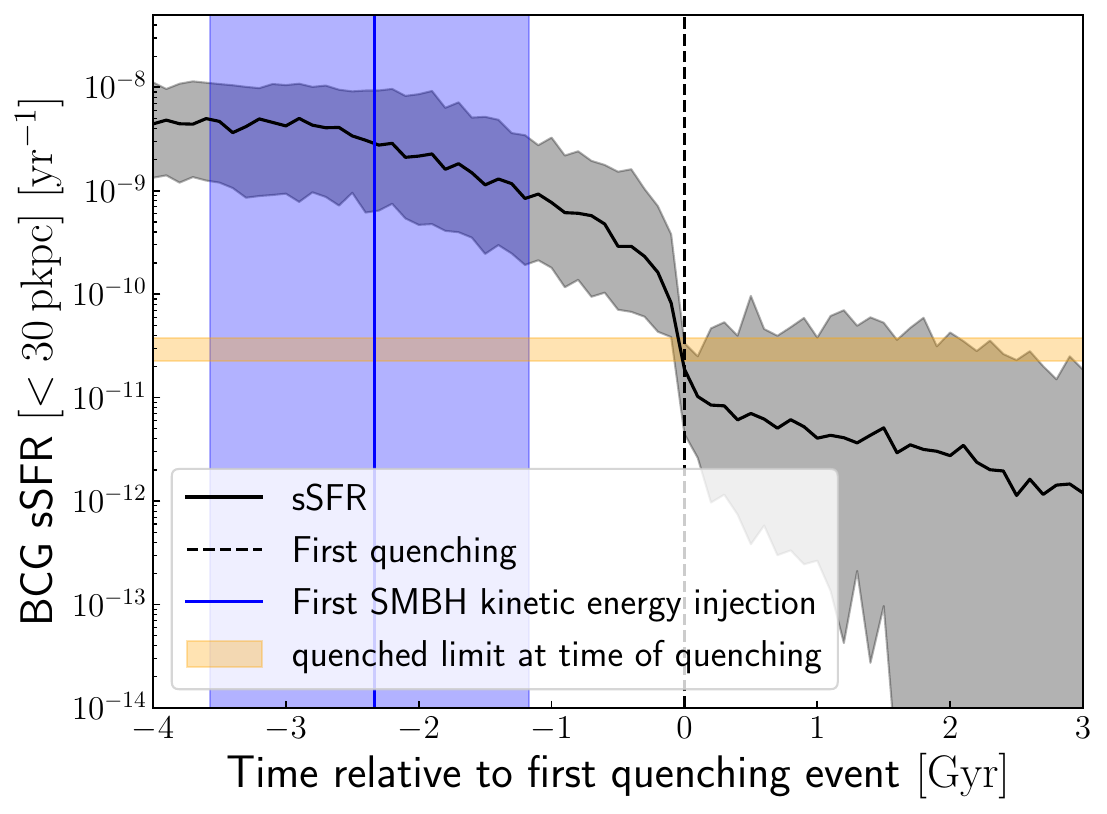}
    \caption{\textbf{The onset of kinetic mode SMBH feedback marks the beginning of the sSFR decline.} Stacked sSFR evolution versus time relative to the first quenching event of all the BCGs in TNG-Cluster. In black, the median sSFR of all BCGs is plotted relative to the time of first quenching, indicated by the black dashed line. The blue vertical line denotes the median time at which the first-ever kinetic energy injection from the SMBH occurs, with the shaded area denoting the 10th-90th percentile. The orange shaded band indicates the 10th-90th percentile range at which the BCGs quench. We see that, simultaneously with the onset of the kinetic mode SMBH feedback $2.3^{+1.2}_{-1.2} \, \mathrm{Gyr}$ before quenching, the sSFR begins to decrease towards the quenched threshold. However, even after switching to kinetic mode SMBH feedback, the BCG still takes a significant amount of time to be quenched.}
    \label{fig:sSFR_before_quenching}
\end{figure}

\subsection{The first SMBH kinetic energy injection does not immediately quench a BCG}

In order to see what consequences SMBH feedback has on the SFRs and star-formation status of the BCGs in our sample, in Fig.~\ref{fig:sSFR_before_quenching} we look at the stacked evolution of the sSFR values, relative to the time of the first quenching event. The black curve indicates the median sSFR, with the grey shaded area representing the 10th-90th percentile. The sSFR is centred on the time of first quenching, represented by the black dashed line. The blue line and shaded area represent the time of the first kinetic energy injection, while the orange shaded band indicates the sSFR range at which the BCGs quench (10th-90th percentile).

The SMBHs deliver their first kinetic energy injection in BCGs about $2.3^{+1.2}_{-1.2} \, \mathrm{Gyr}$ before the first quenching event and continuously transition between the more effective kinetic mode SMBH feedback and the less effective thermal mode SMBH feedback after this point. Around this time, we see a downturn in the sSFR, though still well above the quenched threshold. The SMBHs continue to grow via accretion until they reach the point at which they stay predominantly in kinetic mode SMBH feedback. At this stage, their mass growth slows down and becomes dominated by mergers with other black holes \citep{Weinberger2018, truong2021}. This transition coincides with the steepening of the sSFR decline towards the quenched threshold.

Importantly, we see that the first kinetic energy injection and the SMBH being predominantly in kinetic mode are not enough to immediately shut down star formation in the BCGs. It takes on average more than $2~\mathrm{Gyr}$ of kinetic energy injections before the SMBH is able to quench the BCGs. This explains why there is a large scatter of SMBH masses at first quenching in Fig.~\ref{fig:quenchingtimes}, with all of them significantly higher than the threshold found for normal galaxies. In the next section, we show how this difference is due to the large gravitational binding energies of BCGs.

\begin{figure*}
    \centering
    \includegraphics[width=\linewidth]{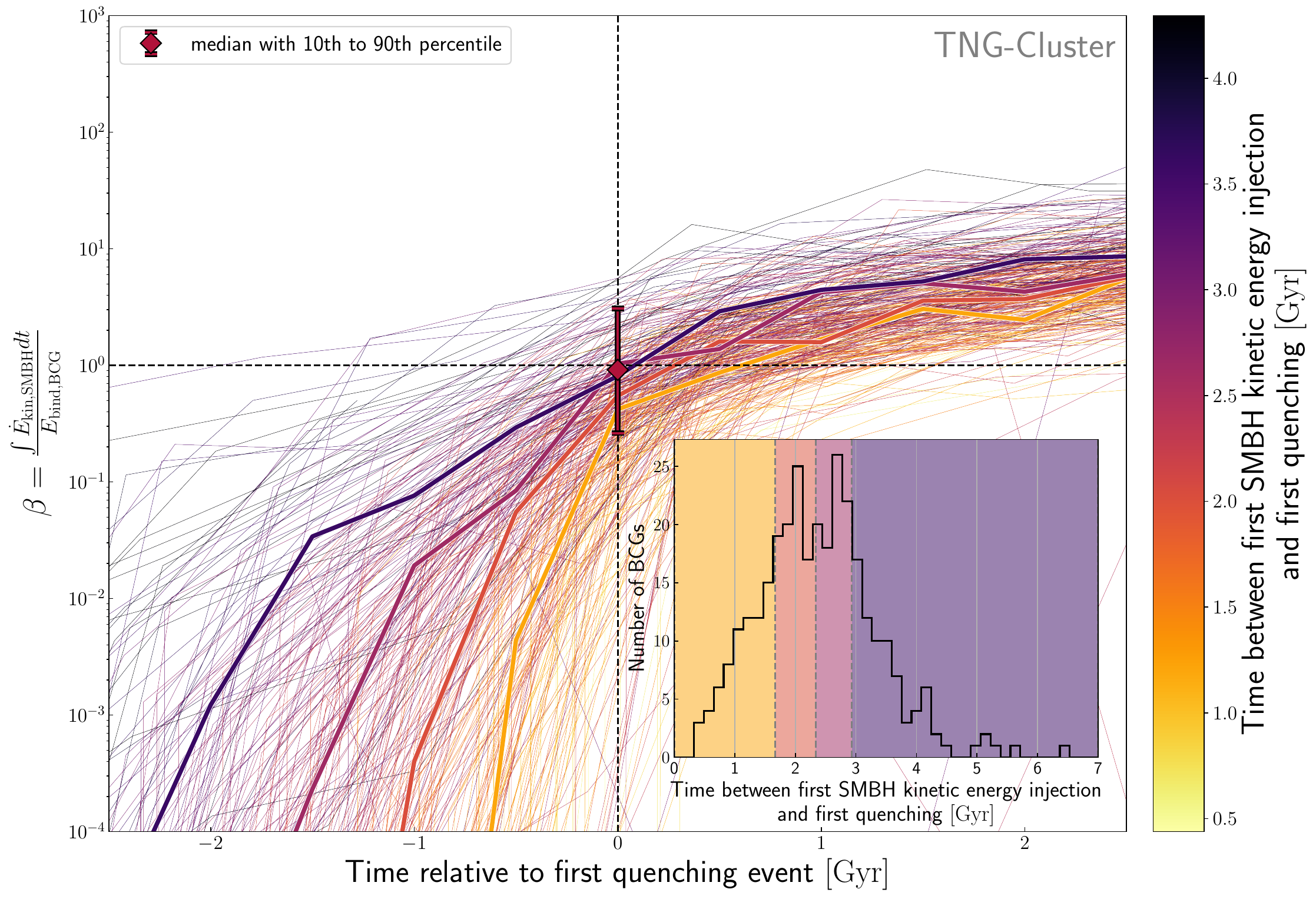}
    \caption{\textbf{According to TNG-Cluster, BCGs quench due to kinetic mode AGN feedback.} Ratio between the cumulative energy injection in kinetic mode and the binding energy of the BCG $\betakin$, plotted versus the time relative to the first quenching event for each individual BCG. The colour-coding represents the time elapsed between the SMBH's first kinetic energy injection and its first quenching event. The thin curves represent each individual BCG, while the thick curves are averages in bins of the time between the first kinetic energy injection and the first quenching event, such that each of the four bins contains 25 per cent of the whole BCG sample. The horizontal dashed line indicates where the binding energy equals the value of the cumulative energy injection, and the vertical dashed line indicates the time of the first quenching event. The ratio grows rapidly before the first quenching event and exceeds 1 around the time of the first quenching for all BCGs. At the time of first quenching, the median ratio is $\betakin = 0.9 ^{+2.2}_{-0.7}$ (10th and 90th percentile), shown by the red square. The inline plot shows a histogram of the distribution of the time evolved between the first kinetic energy injection of SMBH and the first quenching event, visualizing the distribution of the colour-coding in the big plot. In median, the first kinetic energy injection happens at $t=(2.3 \pm 1.0) \, \mathrm{Gyr}$ (standard deviation) before the first quenching event. The distribution shows that the time of the first kinetic energy injection is correlated to the time of first quenching, but this first injection does not instantaneously quench the galaxy. It takes a lot of time and many more energy injections to build up enough energy to shut down star formation.}
    \label{fig:bindingenergy}
\end{figure*}

\subsection{Injected energy equals binding energy at time of first quenching}

To better understand the time delay between the onset of the highly effective SMBH feedback mode and first quenching, we look at the ratio between the cumulative energy injection by the SMBH relative to the binding energy of the BCG, 
\begin{gather}
    \betakin = \frac{\int \Dot{E}_{\mathrm{kin, SMBH}}dt}{E_{\mathrm{bind, BCG}}}
\end{gather}
to find out how the AGN may affect the gas in the BCG around the time of the first quenching. We argue that the AGN might have to inject enough energy to unbind the gas belonging to the BCG before the galaxy can quench. Before that threshold is reached, gas can still effectively fall onto the galaxy and ignite new star formation.

Fig.~\ref{fig:bindingenergy} shows the ratio of injected energy in the kinetic mode to the BCG binding energy relative to the time of first quenching. The vertical dashed line is the time of the first quenching. On the y-axis, the ratio $\betakin$ is plotted logarithmically with a dashed line at $\betakin = 1$, where the injected energy equals the binding energy.
Each thin curve represents one individual BCG, with the colour showing the time elapsed between its first kinetic mode energy injection and the first quenching event.

We find that first quenching happens when $\betakin = 0.9 ^{+2.2}_{-0.7}$, after which it continues to grow. There is a large scatter between the individual BCGs; however, the median value at $t=0$ implies that at the time of quenching, the injected kinetic energy of the AGN equals the binding energy of the BCG, injecting enough momentum into the gas cells of the BCG to remove them, which results in shutting down the star formation. This aligns well with the lack of gas we saw in the gas maps of star-forming and quenched galaxies in Fig.~\ref{fig:gasmaps}. The wide scatter shows that the removal of gas and the quenching do not have to perfectly align in time.

The thick curves are the medians in bins of the time between the first SMBH kinetic energy injection and the quenching and correspond to the coloured regions in the inset histogram. We find that the time over which the energy got injected is weakly correlated with the value of $\betakin$ at the time of first quenching. BCGs which had more gentle energy injections over a longer time before quenching reach, on average, $\betakin = 1$. However, BCGs that had a very rapid buildup of energy quench at $\betakin<1$, indicating that a more violent injection of energy over a short time period can quench BCGs before enough energy is injected to fully unbind the gas.

To put it in a nutshell, from the scatter plots in Fig.~\ref{fig:quenchingtimes} and the histogram in Fig.~\ref{fig:bindingenergy}, we can conclude that neither the first ever energy injection, nor the point in time after which the SMBHs are predominantly in kinetic feedback mode, is when the BCGs quench, but that this happens even later in time. The quenching happens simultaneously with the cumulative injected energy overtaking the binding energy of the BCG, meaning that the gas gets expelled from the BCG, which takes away the fuel for star formation and therefore quenches the galaxy.


\section{Discussion}
\label{sec:discussion}

\begin{figure*}
    \centering
    \includegraphics[width=0.8\linewidth]{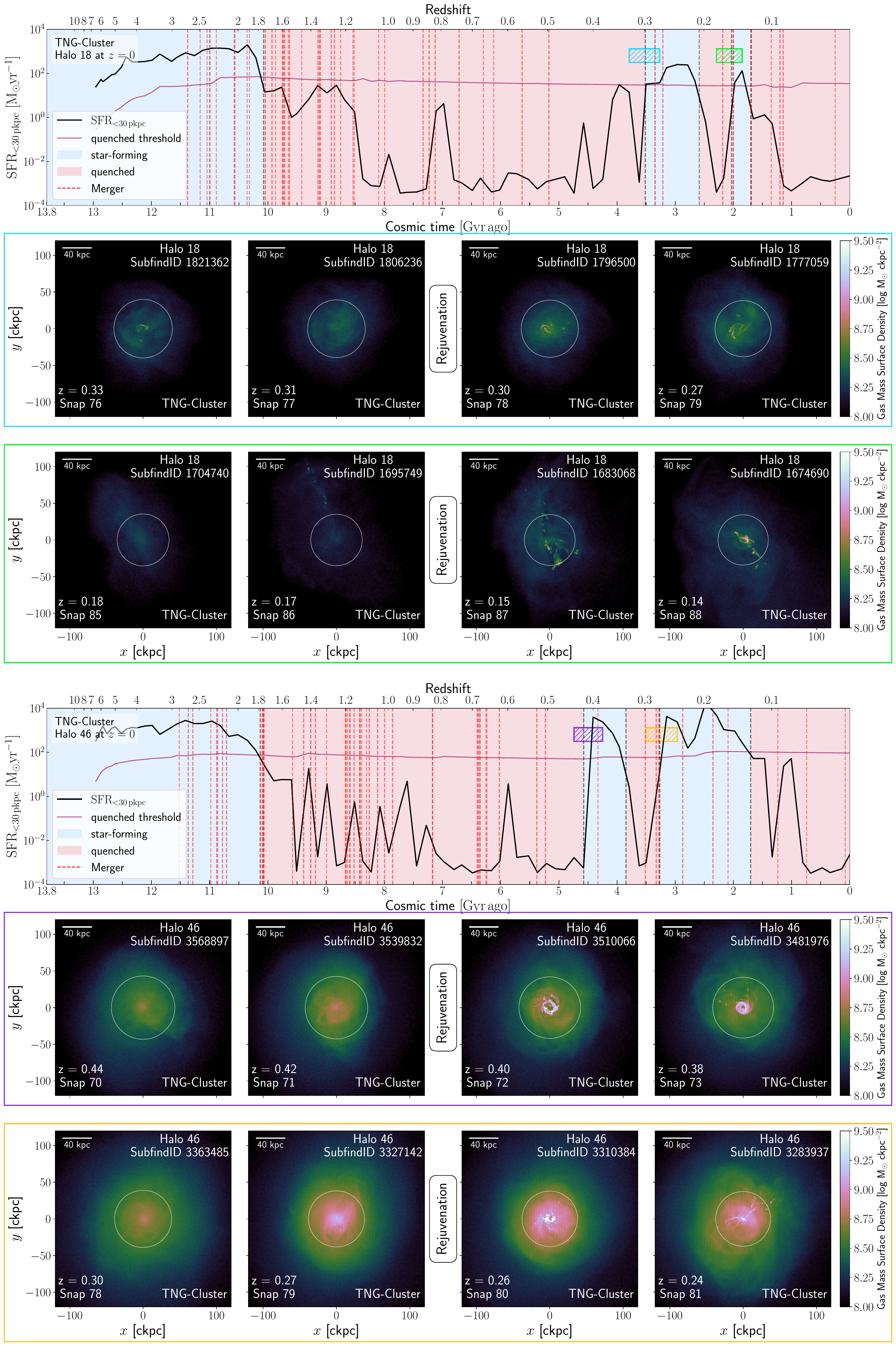}
    \caption{\textbf{Are merger events triggering rejuvenation of the BCGs in TNG-Cluster?} Time sequences of gas column density maps of two selected BCGs to illustrate the behavior around rejuvenation events. The first row shows the star-formation history, similarly to Fig.~\ref{fig:sfr}, with the SFR in black, the quenched threshold in magenta, and the periods of quenching (star formation) in red (blue). The red dotted lines are the times when merger events happened \citep[according to][]{Lehle2025}. The two following rows show the gas mass surface density maps of the BCGs, two snapshots before and two after the rejuvenation event. We display the times of the shown sequences in the star-formation history plot, in the corresponding colour. The white circle denotes the radius of $30 \, \mathrm{pkpc}$. Some rejuvenation events are preceded by one or several merger events, which, in the case of halo 18, are also visible in the maps (green box). In contrast, others have not had any merger events in recent times.}
    \label{fig:rejuvenation_maps}
\end{figure*}
We have characterized the star-forming past of BCGs according to TNG-Cluster, seeing when they first quench, how often they rejuvenate, the relation to stellar, halo, and SMBH mass, and AGN feedback as the driver of first quenching. In this section, we discuss possible mechanisms for rejuvenation, in particular their relation to gas-rich mergers and cool core transformations.

\subsection{Gas-rich mergers can lead to rejuvenation}
We have shown that, despite being mostly quenched today, BCGs rejuvenate their star formation on average twice during their lifetime. Since star formation ceases due to the removal of dense gas in the galaxy, a renewed supply of cold gas can reignite it. A possible mechanism for that is gas-rich merger events, which cause dynamical instabilities that in turn lead to gas inflows into the inner regions of the galaxy, where the gas cools, and a burst of star formation is triggered  \citep[e.g.][]{hernquist1989, Mihos1996, di_matteo2005, webb2015, Pawlik2018}.

To explore how merger events are connected to rejuvenation in BCGs, we show four time sequences of gas density maps for two selected haloes in Fig.~\ref{fig:rejuvenation_maps}: halo 18 (upper panel) and halo 46 (bottom panel). Both haloes have similar halo masses throughout their evolution, with present-day values of $M_{\rm 500c} = 10^{15.2} \, \Msun$ and $10^{15.1} \, \Msun$, respectively. The first row in each panel shows the star-formation history, similar to Fig.~\ref{fig:sfr}, with the black curve representing the SFR and the magenta curve the quenched limit through cosmic time. In the blue shaded periods, the BCG is star forming, while in the red periods, it is quenched. The red dotted lines mark all merger events that the halo is undergoing, according to \citet{Lehle2025}. They are defined as the point in time when the secondary halo crosses $0.5 r_{\rm500c}$ of the primary halo, considering only secondary haloes with a dark matter mass $\rm{M}_{\rm dm} > 10^{11} \, \Msun$.

Below the SFR panels, we show time sequences of gas surface density maps around rejuvenation episodes, corresponding to the points in time marked by the coloured boxes in the star-formation history plot. The rejuvenation events occur between the first and last two frames for halo 18 at $z=0.30$ and $z=0.16$, and for halo 46 at $z=0.42$ and $z=0.27$. The white circle denotes the radius ($30 \, \mathrm{pkpc}$) used for our measurements of SFR.

In all four time series, the gas density slowly increases in the period prior to rejuvenation. Disc features, especially in halo 70, are clearly visible as soon as rejuvenation occurs. If we consider the SFR evolution plot at the times of rejuvenation, we see that in both halo 18 and halo 46, the first rejuvenation event has no mergers in the Gyr prior to rejuvenation, whereas the second rejuvenation episode occurs shortly after one or more merger events. In halo 18, we can speculate that the small spots of high gas density outside the halo core correspond to gaseous inflows driven by the two merger events at that time. These gas inflows are likely the trigger for the burst of star formation between $z=0.16$ and $z=0.13$. In halo 46, despite having two mergers right before quenching at $z = 0.27$, the mergers are not visible in the gas density map, though the morphology shows signs of disturbance at $z=0.24$ after rejuvenation. 

The overall gas densities in the two haloes are also reflected in the levels of SFR reached in the starbursts. In halo 18 the SFR does not surpass $10^3 \, \Msun\mathrm{yr}^{-1}$ and the gas densities in the center of the BCG are low, while halo 46 reaches SFR values of more than $10^4 \, \Msun\mathrm{yr}^{-1}$ while having higher gas densities in the center.

On a population basis, we find that almost all starbursts in the BCGs of the TNG-Cluster have at least one merger event within the Gyr preceding quenching. However, mergers are frequent, and, more often than not, they are not directly followed by a starburst. Furthermore, as we showed in Fig.~\ref{fig:rejuvenation_maps}, there are also rejuvenation events that are not preceded by a merger event.

From this analysis, it is not possible to find a clear correlation between halo mergers and star formation spiking up. 
Furthermore, the merger catalogues in \citet{Lehle2025} consider merger events on a halo scale. However, much smaller radii are relevant for rejuvenations in BCGs, and not every halo merger will have an instantaneous effect on the gas in the BCG at the halo centre.
Therefore, to gain further insight, additional parameters, such as the gas-richness of the merger, should also be considered.

\subsection{Connection to cool core transformations}
In \citet{Nelson2024}, we showed that, in TNG-Cluster, all BCGs that are star forming at $z=0$ are strong cool cores. This indicates that the presence of a cool core in galaxy clusters is related to star formation. We find a similar result when comparing with the cool gas mass instead of star-formation rate. This provides further evidence that the periods when a BCG is star forming might be highly correlated with the periods when it hosts cool or dense gas and, in turn, when it is a cool core.

Furthermore, in Lehle et al. (2026, in prep.), we show that the amount of cool gas depends on the strength of the cool core measure -- stronger cool cores host more cool gas. However, it also depends on the time since the last transformation to a cool core. When a cluster transforms to lower core entropy, i.e., in the direction of being a cool core, it takes some time to build a cool gas reservoir. This might also imply that the times when a BCG gets quenched or resumes star formation might be correlated with the cluster core transformations. Thus, the results and interpretations, especially those related to the drivers of transformations from \citet{Lehle2025}, might be applicable to the results presented here and vice versa. There, we found that cool core transformations are driven by a combination of AGN feedback and mergers, with the two often difficult to separate. Mergers trigger AGN feedback, and it remains unclear which mechanism is ultimately responsible for the removal of gas. This is in line with our results in this work, which indicate that the integrated AGN feedback energy determines the timing of first quenching. Further quenching episodes might be triggered by a combination of mergers and AGN feedback.


\section{Summary and Conclusions}
\label{sec:conclusions}

Brightest cluster galaxies (BCGs) are rare but interesting objects, typically occupying the center of the most massive gravitationally-bound structures in the Universe. Studying them can help us gain deeper insight into various astrophysical and cosmological problems, including how galaxies regulate their star formation and how most of them halt the formation of new stars.

In this paper, we have used the 352 zoom-in simulations of TNG-Cluster, returning galaxy clusters with $M_{\rm 200c} = 10^{14.3-15.4} \, \text{M}_\odot$ at $z=0$, to get insight into the star-forming past of BCGs.

We have evaluated the fraction of quenched BCGs and their gas maps at $z=0$ and quantified the typical timing of the quenching and rejuvenation episodes. We then focused on the first quenching event of the BCGs and investigated the role of AGN feedback. 

We extend previous evidence for kinetic-mode SMBH feedback as the driver of galaxy quenching in IllustrisTNG to the regime of BCGs, showing that the same mechanism governs the evolution of the most massive galaxies. However, star-formation quenching occurs with a $2~\mathrm{Gyr}$ delay after the onset of AGN feedback. In more detail, our results according to TNG-Cluster can be summarized as follows.

\begin{itemize}
    \item At $z=0$, a total of 93 per cent (329) of the BCGs in TNG-Cluster are quiescent, while the remaining 23 BCGs are star forming (Fig.~\ref{fig:sfr} top panel). The star-forming galaxies have high gas densities in their inner regions and show clear disc-like morphologies, while the quenched galaxies have lower gas densities (Fig.~\ref{fig:gasmaps}).
    \item According to TNG-Cluster, BCGs undergo typically between 1 and 3 rejuvenation episodes in their evolution (Fig.~\ref{fig:sfr}, top). The star-formation histories of different BCGs are diverse and show significant variations from one BCG to another (Fig.~\ref{fig:sfr}, bottom panels).
    \item The fraction of permanently-quenched BCGs grows at $z<3$ monotonically until today (Fig.~\ref{fig:sfr} inline plot).
    \item Rejuvenation episodes last on average $500~\mathrm{Myr}$, and become less frequent with cosmic time. Their duration is roughly independent of redshift (Fig.~\ref{fig:rejuvenation_vs_duration}).
    \item The first quenching event happens on average at a cosmic time $9.9^{+0.9}_{-1.4} \, \mathrm{Gyr}$ ago (redshift $z=1.7^{+0.5}_{-0.5}$). The last quenching shows an almost flat distribution from $z \sim 2$ to today, at an average age of $7.0^{+2.8}_{-5.0} \, \mathrm{Gyr}$ ago (Fig.~\ref{fig:quenchingtimes}).
    \item The time of both first and last quenching does not correlate with the halo mass $\Mcrit$ at $z=0$. The first quenching happens at halo masses $\log(\Mcrit/\Msun) = 13.8 ^{+0.4}_{-0.5}$, BCG stellar masses $\log(\Mstar/\Msun) = 11.9 ^{+0.3}_{-0.4}$ and SMBH masses $\log(\MSMBH/\Msun) = 8.9 ^{+0.4}_{-0.2}$, showing that there is a large spread in halo, BCG and SMBH masses at first quenching. The SMBH mass at quenching is almost an order of magnitude higher than the mass at which it transitioned to kinetic mode SMBH feedback.
    \item The last quenching happens at halo masses $\log(\Mcrit/\Msun) = 14.2 ^{+0.5}_{-0.7}$, stellar masses $\log(\Mstar/\Msun) =  12.0 ^{+0.3}_{-0.3}$ and SMBH masses $\log(\MSMBH/\Msun) = 9.3 ^{+0.5}_{-0.5}$ and has a large scatter.
    \item The first quenching happens on average $2.3^{+1.2}_{-1.2} \, \mathrm{Gyr}$ after the first kinetic energy injection, indicating that the first kinetic energy injections are not sufficient to immediately quench the BCGs (Fig.~\ref{fig:sSFR_before_quenching}).
    \item We find that the first quenching happens when the ratio between the cumulative AGN energy injection and the BCG binding energy exceeds $\betakin = 0.9 ^{+2.2}_{-0.7}$, which is the point when the injected energy becomes larger than the binding energy of the BCG and therefore the gas gets expelled from the BCG (Fig.~\ref{fig:bindingenergy}).
\end{itemize}

Overall, our analysis shows that the quenching of the most massive central galaxies in the IllustrisTNG simulations is not a rapid transition tied to a single event or mass threshold, but rather a complex and extended process. In accordance with previous findings, we identify the kinetic-mode AGN feedback to be the primary driver of quenching in TNG-Cluster by expelling the gas from the galaxy cores. However, its impact does not immediately lead to quiescence; rather, it develops over several billion years, with significant object-to-object variation. The diversity in star-formation histories, including multiple rejuvenation episodes, demonstrates that additional processes, such as mergers, also contribute to modulating star formation.

Given the observational challenges of probing quenching mechanisms at high redshift, cosmological simulations are an essential tool for advancing our understanding. Continued efforts to improve AGN feedback models will be crucial for building a more complete picture of galaxy evolution in the most massive structures of our Universe.


\section*{Acknowledgements}

SG, JB, and AP acknowledge funding from the European Union (ERC, COSMIC-KEY, 101087822, PI: Pillepich). KL acknowledges funding from the Hector Fellow Academy through a Research Career Development Award. DN acknowledges funding from the Deutsche Forschungsgemeinschaft (DFG) through an Emmy Noether Research Group (grant number NE 2441/1-1).
SG thanks Jehanne Delhomelle for her contribution of the code used in generating the gas density maps.

\section*{Data Availability}

The measurements of the simulation data performed for this analysis will be made publicly available upon acceptance for publication. All the IllustrisTNG simulations, including TNG-Cluster, are publicly available and accessible at \url{www.tng-project.org/data}, as described in \citet{Nelson2019b}.


\bibliographystyle{mnras}
\bibliography{literature}

@ARTICLE{Nelson2024,
       author = {{Nelson}, Dylan and {Pillepich}, Annalisa and {Ayromlou}, Mohammadreza and {Lee}, Wonki and {Lehle}, Katrin and {Rohr}, Eric and {Truong}, Nhut},
        title = "{Introducing the TNG-Cluster simulation: Overview and the physical properties of the gaseous intracluster medium}",
      journal = {\aap},
         year = 2024,
        month = jun,
       volume = {686},
          eid = {A157},
        pages = {A157},
          doi = {10.1051/0004-6361/202348608},
archivePrefix = {arXiv},
       eprint = {2311.06338},
 primaryClass = {astro-ph.GA},
       adsurl = {https://ui.adsabs.harvard.edu/abs/2024A&A...686A.157N}
}

@ARTICLE{Nelson2019b,
       author = {{Nelson}, Dylan and {Springel}, Volker and {Pillepich}, Annalisa and {Rodriguez-Gomez}, Vicente and {Torrey}, Paul and {Genel}, Shy and {Vogelsberger}, Mark and {Pakmor}, Ruediger and {Marinacci}, Federico and {Weinberger}, Rainer and {Kelley}, Luke and {Lovell}, Mark and {Diemer}, Benedikt and {Hernquist}, Lars},
        title = "{The IllustrisTNG simulations: public data release}",
      journal = {Computational Astrophysics and Cosmology},
         year = 2019,
        month = may,
       volume = {6},
       number = {1},
          eid = {2},
        pages = {2},
          doi = {10.1186/s40668-019-0028-x},
archivePrefix = {arXiv},
       eprint = {1812.05609},
 primaryClass = {astro-ph.GA},
       adsurl = {https://ui.adsabs.harvard.edu/abs/2019ComAC...6....2N}
}

@ARTICLE{Staffehl2025,
       author = {{Staffehl}, Milan and {Nelson}, Dylan and {Ayromlou}, Mohammadreza and {Rohr}, Eric and {Pillepich}, Annalisa},
        title = "{The abundance and origin of cool gas in galaxy clusters in the TNG-Cluster simulation}",
      journal = {\aap},
         year = 2025,
        month = nov,
       volume = {703},
          eid = {A33},
        pages = {A33},
          doi = {10.1051/0004-6361/202554373},
archivePrefix = {arXiv},
       eprint = {2503.01960},
 primaryClass = {astro-ph.GA},
       adsurl = {https://ui.adsabs.harvard.edu/abs/2025A&A...703A..33S}
}

@ARTICLE{Donnari2021,
       author = {{Donnari}, Martina and {Pillepich}, Annalisa and {Joshi}, Gandhali D. and {Nelson}, Dylan and {Genel}, Shy and {Marinacci}, Federico and {Rodriguez-Gomez}, Vicente and {Pakmor}, R{\"u}diger and {Torrey}, Paul and {Vogelsberger}, Mark and {Hernquist}, Lars},
        title = "{Quenched fractions in the IllustrisTNG simulations: the roles of AGN feedback, environment, and pre-processing}",
      journal = {\mnras},
         year = 2021,
        month = jan,
       volume = {500},
       number = {3},
        pages = {4004-4024},
          doi = {10.1093/mnras/staa3006},
archivePrefix = {arXiv},
       eprint = {2008.00005},
 primaryClass = {astro-ph.GA},
       adsurl = {https://ui.adsabs.harvard.edu/abs/2021MNRAS.500.4004D}
}

@ARTICLE{Donnari2019,
       author = {{Donnari}, Martina and {Pillepich}, Annalisa and {Nelson}, Dylan and {Vogelsberger}, Mark and {Genel}, Shy and {Weinberger}, Rainer and {Marinacci}, Federico and {Springel}, Volker and {Hernquist}, Lars},
        title = "{The star formation activity of IllustrisTNG galaxies: main sequence, UVJ diagram, quenched fractions, and systematics}",
      journal = {\mnras},
         year = 2019,
        month = jun,
       volume = {485},
       number = {4},
        pages = {4817-4840},
          doi = {10.1093/mnras/stz712},
archivePrefix = {arXiv},
       eprint = {1812.07584},
 primaryClass = {astro-ph.GA},
       adsurl = {https://ui.adsabs.harvard.edu/abs/2019MNRAS.485.4817D}
}

@ARTICLE{Matthews1964,
       author = {{Matthews}, Thomas A. and {Morgan}, William W. and {Schmidt}, Maarten},
        title = {A Discussion of Galaxies Indentified with Radio Sources.},
      journal = {APJ},
         year = {1964},
        month = {jul},
       volume = {140},
        pages = {35},
          doi = {10.1086/147890},
       adsurl = {https://ui.adsabs.harvard.edu/abs/1964ApJ...140...35M}
}

@ARTICLE{Dubinski1998,
       author = {{Dubinski}, John},
        title = "{The Origin of the Brightest Cluster Galaxies}",
      journal = {\apj},
         year = 1998,
        month = jul,
       volume = {502},
       number = {1},
        pages = {141-149},
          doi = {10.1086/305901},
archivePrefix = {arXiv},
       eprint = {astro-ph/9709102},
 primaryClass = {astro-ph},
       adsurl = {https://ui.adsabs.harvard.edu/abs/1998ApJ...502..141D}
}

@ARTICLE{wetzel2014galaxy,
       author = {{Wetzel}, Andrew R. and {Tinker}, Jeremy L. and {Conroy}, Charlie and {van den Bosch}, Frank C.},
        title = "{Galaxy evolution near groups and clusters: ejected satellites and the spatial extent of environmental quenching}",
      journal = {\mnras},
         year = 2014,
        month = apr,
       volume = {439},
       number = {3},
        pages = {2687-2700},
          doi = {10.1093/mnras/stu122},
archivePrefix = {arXiv},
       eprint = {1303.7231},
 primaryClass = {astro-ph.CO},
       adsurl = {https://ui.adsabs.harvard.edu/abs/2014MNRAS.439.2687W}
}

@ARTICLE{Springel2005a,
       author = {{Springel}, Volker and {Di Matteo}, Tiziana and {Hernquist}, Lars},
        title = "{Modelling feedback from stars and black holes in galaxy mergers}",
      journal = {\mnras},
         year = 2005,
        month = aug,
       volume = {361},
       number = {3},
        pages = {776-794},
          doi = {10.1111/j.1365-2966.2005.09238.x},
archivePrefix = {arXiv},
       eprint = {astro-ph/0411108},
 primaryClass = {astro-ph},
       adsurl = {https://ui.adsabs.harvard.edu/abs/2005MNRAS.361..776S}
}

@ARTICLE{Chu2021,
       author = {{Chu}, A. and {Durret}, F. and {M{\'a}rquez}, I.},
        title = "{Physical properties of brightest cluster galaxies up to redshift 1.80 based on HST data}",
      journal = {\aap},
         year = 2021,
        month = may,
       volume = {649},
          eid = {A42},
        pages = {A42},
          doi = {10.1051/0004-6361/202040245},
archivePrefix = {arXiv},
       eprint = {2102.01557},
 primaryClass = {astro-ph.GA},
       adsurl = {https://ui.adsabs.harvard.edu/abs/2021A&A...649A..42C}
}

@ARTICLE{Weinberger2018,
       author = {{Weinberger}, Rainer and {Springel}, Volker and {Pakmor}, R{\"u}diger and {Nelson}, Dylan and {Genel}, Shy and {Pillepich}, Annalisa and {Vogelsberger}, Mark and {Marinacci}, Federico and {Naiman}, Jill and {Torrey}, Paul and {Hernquist}, Lars},
        title = "{Supermassive black holes and their feedback effects in the IllustrisTNG simulation}",
      journal = {\mnras},
         year = 2018,
        month = sep,
       volume = {479},
       number = {3},
        pages = {4056-4072},
          doi = {10.1093/mnras/sty1733},
archivePrefix = {arXiv},
       eprint = {1710.04659},
 primaryClass = {astro-ph.GA},
       adsurl = {https://ui.adsabs.harvard.edu/abs/2018MNRAS.479.4056W}
}

@ARTICLE{Corasaniti2021,
       author = {{Corasaniti}, Pier-Stefano and {Sereno}, Mauro and {Ettori}, Stefano},
        title = "{Cosmological Constraints from Galaxy Cluster Sparsity, Cluster Gas Mass Fraction, and Baryon Acoustic Oscillation Data}",
      journal = {\apj},
         year = 2021,
        month = apr,
       volume = {911},
       number = {2},
          eid = {82},
        pages = {82},
          doi = {10.3847/1538-4357/abe9a4},
archivePrefix = {arXiv},
       eprint = {2103.03283},
 primaryClass = {astro-ph.CO},
       adsurl = {https://ui.adsabs.harvard.edu/abs/2021ApJ...911...82C}
}

@ARTICLE{Hektor_2012,
       author = {{Hektor}, A. and {Raidal}, M. and {Tempel}, E.},
        title = "{Evidence for Indirect Detection of Dark Matter from Galaxy Clusters in Fermi {\ensuremath{\gamma}}-Ray Data}",
      journal = {\apjl},
         year = 2013,
        month = jan,
       volume = {762},
       number = {2},
          eid = {L22},
        pages = {L22},
          doi = {10.1088/2041-8205/762/2/L22},
archivePrefix = {arXiv},
       eprint = {1207.4466},
 primaryClass = {astro-ph.HE},
       adsurl = {https://ui.adsabs.harvard.edu/abs/2013ApJ...762L..22H}
}

@ARTICLE{Madau2014,
       author = {{Madau}, Piero and {Dickinson}, Mark},
        title = "{Cosmic Star-Formation History}",
      journal = {\araa},
         year = 2014,
        month = aug,
       volume = {52},
        pages = {415-486},
          doi = {10.1146/annurev-astro-081811-125615},
archivePrefix = {arXiv},
       eprint = {1403.0007},
 primaryClass = {astro-ph.CO},
       adsurl = {https://ui.adsabs.harvard.edu/abs/2014ARA&A..52..415M}
}

@ARTICLE{Marinacci2018,
       author = {{Marinacci}, Federico and {Vogelsberger}, Mark and {Pakmor}, R{\"u}diger and {Torrey}, Paul and {Springel}, Volker and {Hernquist}, Lars and {Nelson}, Dylan and {Weinberger}, Rainer and {Pillepich}, Annalisa and {Naiman}, Jill and {Genel}, Shy},
        title = "{First results from the IllustrisTNG simulations: radio haloes and magnetic fields}",
      journal = {\mnras},
         year = 2018,
        month = nov,
       volume = {480},
       number = {4},
        pages = {5113-5139},
          doi = {10.1093/mnras/sty2206},
archivePrefix = {arXiv},
       eprint = {1707.03396},
 primaryClass = {astro-ph.CO},
       adsurl = {https://ui.adsabs.harvard.edu/abs/2018MNRAS.480.5113M}
}

@ARTICLE{naiman2018,
       author = {{Naiman}, Jill P. and {Pillepich}, Annalisa and {Springel}, Volker and {Ramirez-Ruiz}, Enrico and {Torrey}, Paul and {Vogelsberger}, Mark and {Pakmor}, R{\"u}diger and {Nelson}, Dylan and {Marinacci}, Federico and {Hernquist}, Lars and {Weinberger}, Rainer and {Genel}, Shy},
        title = "{First results from the IllustrisTNG simulations: a tale of two elements - chemical evolution of magnesium and europium}",
      journal = {\mnras},
         year = 2018,
        month = jun,
       volume = {477},
       number = {1},
        pages = {1206-1224},
          doi = {10.1093/mnras/sty618},
archivePrefix = {arXiv},
       eprint = {1707.03401},
 primaryClass = {astro-ph.GA},
       adsurl = {https://ui.adsabs.harvard.edu/abs/2018MNRAS.477.1206N}
}

@ARTICLE{Springel2010,
       author = {{Springel}, Volker},
        title = "{E pur si muove: Galilean-invariant cosmological hydrodynamical simulations on a moving mesh}",
      journal = {\mnras},
         year = 2010,
        month = jan,
       volume = {401},
       number = {2},
        pages = {791-851},
          doi = {10.1111/j.1365-2966.2009.15715.x},
archivePrefix = {arXiv},
       eprint = {0901.4107},
 primaryClass = {astro-ph.CO},
       adsurl = {https://ui.adsabs.harvard.edu/abs/2010MNRAS.401..791S}
}

@ARTICLE{Rohr2025,
       author = {{Rohr}, Eric and {Pillepich}, Annalisa and {Nelson}, Dylan and {Ayromlou}, Mohammadreza and {P{\'e}roux}, C{\'e}line and {Zinger}, Elad},
        title = "{The cooler past of the intracluster medium in TNG-cluster}",
      journal = {\mnras},
         year = 2025,
        month = jan,
       volume = {536},
       number = {2},
        pages = {1226-1250},
          doi = {10.1093/mnras/stae2536},
archivePrefix = {arXiv},
       eprint = {2410.19900},
 primaryClass = {astro-ph.GA},
       adsurl = {https://ui.adsabs.harvard.edu/abs/2025MNRAS.536.1226R}
}

@ARTICLE{Planck2016,
       author = {{Planck Collaboration} and {Ade}, P.~A.~R. and {Aghanim}, N. and {Arnaud}, M. and {Ashdown}, M. and {Aumont}, J. and {Baccigalupi}, C. and {Banday}, A.~J. and {Barreiro}, R.~B. and {Bartlett}, J.~G. and {Bartolo}, N. and {Battaner}, E. and {Battye}, R. and {Benabed}, K. and {Beno{\^\i}t}, A. and {Benoit-L{\'e}vy}, A. and {Bernard}, J.-P. and {Bersanelli}, M. and {Bielewicz}, P. and {Bock}, J.~J. and {Bonaldi}, A. and {Bonavera}, L. and {Bond}, J.~R. and {Borrill}, J. and {Bouchet}, F.~R. and {Boulanger}, F. and {Bucher}, M. and {Burigana}, C. and {Butler}, R.~C. and {Calabrese}, E. and {Cardoso}, J.-F. and {Catalano}, A. and {Challinor}, A. and {Chamballu}, A. and {Chary}, R.-R. and {Chiang}, H.~C. and {Chluba}, J. and {Christensen}, P.~R. and {Church}, S. and {Clements}, D.~L. and {Colombi}, S. and {Colombo}, L.~P.~L. and {Combet}, C. and {Coulais}, A. and {Crill}, B.~P. and {Curto}, A. and {Cuttaia}, F. and {Danese}, L. and {Davies}, R.~D. and {Davis}, R.~J. and {de Bernardis}, P. and {de Rosa}, A. and {de Zotti}, G. and {Delabrouille}, J. and {D{\'e}sert}, F.-X. and {Di Valentino}, E. and {Dickinson}, C. and {Diego}, J.~M. and {Dolag}, K. and {Dole}, H. and {Donzelli}, S. and {Dor{\'e}}, O. and {Douspis}, M. and {Ducout}, A. and {Dunkley}, J. and {Dupac}, X. and {Efstathiou}, G. and {Elsner}, F. and {En{\ss}lin}, T.~A. and {Eriksen}, H.~K. and {Farhang}, M. and {Fergusson}, J. and {Finelli}, F. and {Forni}, O. and {Frailis}, M. and {Fraisse}, A.~A. and {Franceschi}, E. and {Frejsel}, A. and {Galeotta}, S. and {Galli}, S. and {Ganga}, K. and {Gauthier}, C. and {Gerbino}, M. and {Ghosh}, T. and {Giard}, M. and {Giraud-H{\'e}raud}, Y. and {Giusarma}, E. and {Gjerl{\o}w}, E. and {Gonz{\'a}lez-Nuevo}, J. and {G{\'o}rski}, K.~M. and {Gratton}, S. and {Gregorio}, A. and {Gruppuso}, A. and {Gudmundsson}, J.~E. and {Hamann}, J. and {Hansen}, F.~K. and {Hanson}, D. and {Harrison}, D.~L. and {Helou}, G. and {Henrot-Versill{\'e}}, S. and {Hern{\'a}ndez-Monteagudo}, C. and {Herranz}, D. and {Hildebrandt}, S.~R. and {Hivon}, E. and {Hobson}, M. and {Holmes}, W.~A. and {Hornstrup}, A. and {Hovest}, W. and {Huang}, Z. and {Huffenberger}, K.~M. and {Hurier}, G. and {Jaffe}, A.~H. and {Jaffe}, T.~R. and {Jones}, W.~C. and {Juvela}, M. and {Keih{\"a}nen}, E. and {Keskitalo}, R. and {Kisner}, T.~S. and {Kneissl}, R. and {Knoche}, J. and {Knox}, L. and {Kunz}, M. and {Kurki-Suonio}, H. and {Lagache}, G. and {L{\"a}hteenm{\"a}ki}, A. and {Lamarre}, J.-M. and {Lasenby}, A. and {Lattanzi}, M. and {Lawrence}, C.~R. and {Leahy}, J.~P. and {Leonardi}, R. and {Lesgourgues}, J. and {Levrier}, F. and {Lewis}, A. and {Liguori}, M. and {Lilje}, P.~B. and {Linden-V{\o}rnle}, M. and {L{\'o}pez-Caniego}, M. and {Lubin}, P.~M. and {Mac{\'\i}as-P{\'e}rez}, J.~F. and {Maggio}, G. and {Maino}, D. and {Mandolesi}, N. and {Mangilli}, A. and {Marchini}, A. and {Maris}, M. and {Martin}, P.~G. and {Martinelli}, M. and {Mart{\'\i}nez-Gonz{\'a}lez}, E. and {Masi}, S. and {Matarrese}, S. and {McGehee}, P. and {Meinhold}, P.~R. and {Melchiorri}, A. and {Melin}, J.-B. and {Mendes}, L. and {Mennella}, A. and {Migliaccio}, M. and {Millea}, M. and {Mitra}, S. and {Miville-Desch{\^e}nes}, M.-A. and {Moneti}, A. and {Montier}, L. and {Morgante}, G. and {Mortlock}, D. and {Moss}, A. and {Munshi}, D. and {Murphy}, J.~A. and {Naselsky}, P. and {Nati}, F. and {Natoli}, P. and {Netterfield}, C.~B. and {N{\o}rgaard-Nielsen}, H.~U. and {Noviello}, F. and {Novikov}, D. and {Novikov}, I. and {Oxborrow}, C.~A. and {Paci}, F. and {Pagano}, L. and {Pajot}, F. and {Paladini}, R. and {Paoletti}, D. and {Partridge}, B. and {Pasian}, F. and {Patanchon}, G. and {Pearson}, T.~J. and {Perdereau}, O. and {Perotto}, L. and {Perrotta}, F. and {Pettorino}, V. and {Piacentini}, F. and {Piat}, M. and {Pierpaoli}, E. and {Pietrobon}, D. and {Plaszczynski}, S. and {Pointecouteau}, E. and {Polenta}, G. and {Popa}, L. and {Pratt}, G.~W. and {Pr{\'e}zeau}, G.},
        title = "{Planck 2015 results. XIII. Cosmological parameters}",
      journal = {\aap},
         year = 2016,
        month = sep,
       volume = {594},
          eid = {A13},
        pages = {A13},
          doi = {10.1051/0004-6361/201525830},
archivePrefix = {arXiv},
       eprint = {1502.01589},
 primaryClass = {astro-ph.CO},
       adsurl = {https://ui.adsabs.harvard.edu/abs/2016A&A...594A..13P}
}

@ARTICLE{Weinberger2017,
       author = {{Weinberger}, Rainer and {Springel}, Volker and {Hernquist}, Lars and {Pillepich}, Annalisa and {Marinacci}, Federico and {Pakmor}, R{\"u}diger and {Nelson}, Dylan and {Genel}, Shy and {Vogelsberger}, Mark and {Naiman}, Jill and {Torrey}, Paul},
        title = "{Simulating galaxy formation with black hole driven thermal and kinetic feedback}",
      journal = {\mnras},
         year = 2017,
        month = mar,
       volume = {465},
       number = {3},
        pages = {3291-3308},
          doi = {10.1093/mnras/stw2944},
archivePrefix = {arXiv},
       eprint = {1607.03486},
 primaryClass = {astro-ph.GA},
       adsurl = {https://ui.adsabs.harvard.edu/abs/2017MNRAS.465.3291W}
}

@ARTICLE{Pillepich2018b,
       author = {{Pillepich}, Annalisa and {Springel}, Volker and {Nelson}, Dylan and {Genel}, Shy and {Naiman}, Jill and {Pakmor}, R{\"u}diger and {Hernquist}, Lars and {Torrey}, Paul and {Vogelsberger}, Mark and {Weinberger}, Rainer and {Marinacci}, Federico},
        title = "{Simulating galaxy formation with the IllustrisTNG model}",
      journal = {\mnras},
         year = 2018,
        month = jan,
       volume = {473},
       number = {3},
        pages = {4077-4106},
          doi = {10.1093/mnras/stx2656},
archivePrefix = {arXiv},
       eprint = {1703.02970},
 primaryClass = {astro-ph.GA},
       adsurl = {https://ui.adsabs.harvard.edu/abs/2018MNRAS.473.4077P}
}

@ARTICLE{Springel2001,
       author = {{Springel}, Volker and {White}, Simon D.~M. and {Tormen}, Giuseppe and {Kauffmann}, Guinevere},
        title = "{Populating a cluster of galaxies - I. Results at z=0}",
      journal = {\mnras},
         year = 2001,
        month = dec,
       volume = {328},
       number = {3},
        pages = {726-750},
          doi = {10.1046/j.1365-8711.2001.04912.x},
archivePrefix = {arXiv},
       eprint = {astro-ph/0012055},
 primaryClass = {astro-ph},
       adsurl = {https://ui.adsabs.harvard.edu/abs/2001MNRAS.328..726S}
}

@ARTICLE{Rodriguez2015,
       author = {{Rodriguez-Gomez}, Vicente and {Genel}, Shy and {Vogelsberger}, Mark and {Sijacki}, Debora and {Pillepich}, Annalisa and {Sales}, Laura V. and {Torrey}, Paul and {Snyder}, Greg and {Nelson}, Dylan and {Springel}, Volker and {Ma}, Chung-Pei and {Hernquist}, Lars},
        title = "{The merger rate of galaxies in the Illustris simulation: a comparison with observations and semi-empirical models}",
      journal = {\mnras},
         year = 2015,
        month = may,
       volume = {449},
       number = {1},
        pages = {49-64},
          doi = {10.1093/mnras/stv264},
archivePrefix = {arXiv},
       eprint = {1502.01339},
 primaryClass = {astro-ph.GA},
       adsurl = {https://ui.adsabs.harvard.edu/abs/2015MNRAS.449...49R}
}

@ARTICLE{Pillepich2018a,
       author = {{Pillepich}, Annalisa and {Nelson}, Dylan and {Hernquist}, Lars and {Springel}, Volker and {Pakmor}, R{\"u}diger and {Torrey}, Paul and {Weinberger}, Rainer and {Genel}, Shy and {Naiman}, Jill P. and {Marinacci}, Federico and {Vogelsberger}, Mark},
        title = "{First results from the IllustrisTNG simulations: the stellar mass content of groups and clusters of galaxies}",
      journal = {\mnras},
         year = 2018,
        month = mar,
       volume = {475},
       number = {1},
        pages = {648-675},
          doi = {10.1093/mnras/stx3112},
archivePrefix = {arXiv},
       eprint = {1707.03406},
 primaryClass = {astro-ph.GA},
       adsurl = {https://ui.adsabs.harvard.edu/abs/2018MNRAS.475..648P}
}

@ARTICLE{Springel2018,
       author = {{Springel}, Volker and {Pakmor}, R{\"u}diger and {Pillepich}, Annalisa and {Weinberger}, Rainer and {Nelson}, Dylan and {Hernquist}, Lars and {Vogelsberger}, Mark and {Genel}, Shy and {Torrey}, Paul and {Marinacci}, Federico and {Naiman}, Jill},
        title = "{First results from the IllustrisTNG simulations: matter and galaxy clustering}",
      journal = {\mnras},
         year = 2018,
        month = mar,
       volume = {475},
       number = {1},
        pages = {676-698},
          doi = {10.1093/mnras/stx3304},
archivePrefix = {arXiv},
       eprint = {1707.03397},
 primaryClass = {astro-ph.GA},
       adsurl = {https://ui.adsabs.harvard.edu/abs/2018MNRAS.475..676S}
}

@ARTICLE{Nelson2018a,
       author = {{Nelson}, Dylan and {Pillepich}, Annalisa and {Springel}, Volker and {Weinberger}, Rainer and {Hernquist}, Lars and {Pakmor}, R{\"u}diger and {Genel}, Shy and {Torrey}, Paul and {Vogelsberger}, Mark and {Kauffmann}, Guinevere and {Marinacci}, Federico and {Naiman}, Jill},
        title = "{First results from the IllustrisTNG simulations: the galaxy colour bimodality}",
      journal = {\mnras},
         year = 2018,
        month = mar,
       volume = {475},
       number = {1},
        pages = {624-647},
          doi = {10.1093/mnras/stx3040},
archivePrefix = {arXiv},
       eprint = {1707.03395},
 primaryClass = {astro-ph.GA},
       adsurl = {https://ui.adsabs.harvard.edu/abs/2018MNRAS.475..624N}
}

@ARTICLE{Davis85,
       author = {{Davis}, M. and {Efstathiou}, G. and {Frenk}, C.~S. and {White}, S.~D.~M.},
        title = "{The evolution of large-scale structure in a universe dominated by cold dark matter}",
      journal = {\apj},
         year = 1985,
        month = may,
       volume = {292},
        pages = {371-394},
          doi = {10.1086/163168},
       adsurl = {https://ui.adsabs.harvard.edu/abs/1985ApJ...292..371D}
}

@ARTICLE{Tacchella2019,
       author = {{Tacchella}, Sandro and {Diemer}, Benedikt and {Hernquist}, Lars and {Genel}, Shy and {Marinacci}, Federico and {Nelson}, Dylan and {Pillepich}, Annalisa and {Rodriguez-Gomez}, Vicente and {Sales}, Laura V. and {Springel}, Volker and {Vogelsberger}, Mark},
        title = "{Morphology and star formation in IllustrisTNG: the build-up of spheroids and discs}",
      journal = {\mnras},
         year = 2019,
        month = aug,
       volume = {487},
       number = {4},
        pages = {5416-5440},
          doi = {10.1093/mnras/stz1657},
archivePrefix = {arXiv},
       eprint = {1904.12860},
 primaryClass = {astro-ph.GA},
       adsurl = {https://ui.adsabs.harvard.edu/abs/2019MNRAS.487.5416T}
}

@ARTICLE{Whitaker2012,
       author = {{Whitaker}, Katherine E. and {van Dokkum}, Pieter G. and {Brammer}, Gabriel and {Franx}, Marijn},
        title = "{The Star Formation Mass Sequence Out to z = 2.5}",
      journal = {\apjl},
         year = 2012,
        month = aug,
       volume = {754},
       number = {2},
          eid = {L29},
        pages = {L29},
          doi = {10.1088/2041-8205/754/2/L29},
archivePrefix = {arXiv},
       eprint = {1205.0547},
 primaryClass = {astro-ph.CO},
       adsurl = {https://ui.adsabs.harvard.edu/abs/2012ApJ...754L..29W}
}

@ARTICLE{Kluge2020,
       author = {{Kluge}, M. and {Neureiter}, B. and {Riffeser}, A. and {Bender}, R. and {Goessl}, C. and {Hopp}, U. and {Schmidt}, M. and {Ries}, C. and {Brosch}, N.},
        title = "{Structure of Brightest Cluster Galaxies and Intracluster Light}",
      journal = {\apjs},
         year = 2020,
        month = apr,
       volume = {247},
       number = {2},
          eid = {43},
        pages = {43},
          doi = {10.3847/1538-4365/ab733b},
archivePrefix = {arXiv},
       eprint = {1908.08544},
 primaryClass = {astro-ph.GA},
       adsurl = {https://ui.adsabs.harvard.edu/abs/2020ApJS..247...43K}
}

@ARTICLE{Bai2014,
       author = {{Bai}, Lei and {Yee}, H.~K.~C. and {Yan}, Renbin and {Lee}, Eve and {Gilbank}, David G. and {Ellingson}, E. and {Barrientos}, L.~F. and {Gladders}, M.~D. and {Hsieh}, B.~C. and {Li}, I.~H.},
        title = "{The Inside-out Growth of the Most Massive Galaxies at 0.3 < z < 0.9}",
      journal = {\apj},
         year = 2014,
        month = jul,
       volume = {789},
       number = {2},
          eid = {134},
        pages = {134},
          doi = {10.1088/0004-637X/789/2/134},
archivePrefix = {arXiv},
       eprint = {1406.4149},
 primaryClass = {astro-ph.GA},
       adsurl = {https://ui.adsabs.harvard.edu/abs/2014ApJ...789..134B}
}

@ARTICLE{Zwicky1951,
       author = {{Zwicky}, F.},
        title = "{The Coma Cluster of Galaxies}",
      journal = {\pasp},
         year = 1951,
        month = apr,
       volume = {63},
       number = {371},
        pages = {61},
          doi = {10.1086/126318},
       adsurl = {https://ui.adsabs.harvard.edu/abs/1951PASP...63...61Z}
}

@ARTICLE{Terrazas2020,
       author = {{Terrazas}, Bryan A. and {Bell}, Eric F. and {Pillepich}, Annalisa and {Nelson}, Dylan and {Somerville}, Rachel S. and {Genel}, Shy and {Weinberger}, Rainer and {Habouzit}, M{\'e}lanie and {Li}, Yuan and {Hernquist}, Lars and {Vogelsberger}, Mark},
        title = "{The relationship between black hole mass and galaxy properties: examining the black hole feedback model in IllustrisTNG}",
      journal = {\mnras},
         year = 2020,
        month = apr,
       volume = {493},
       number = {2},
        pages = {1888-1906},
          doi = {10.1093/mnras/staa374},
archivePrefix = {arXiv},
       eprint = {1906.02747},
 primaryClass = {astro-ph.GA},
       adsurl = {https://ui.adsabs.harvard.edu/abs/2020MNRAS.493.1888T}
}

@ARTICLE{Rupke2017,
       author = {{Rupke}, David S.~N. and {G{\"u}ltekin}, Kayhan and {Veilleux}, Sylvain},
        title = "{Quasar-mode Feedback in Nearby Type 1 Quasars: Ubiquitous Kiloparsec-scale Outflows and Correlations with Black Hole Properties}",
      journal = {\apj},
         year = 2017,
        month = nov,
       volume = {850},
       number = {1},
          eid = {40},
        pages = {40},
          doi = {10.3847/1538-4357/aa94d1},
archivePrefix = {arXiv},
       eprint = {1708.05139},
 primaryClass = {astro-ph.GA},
       adsurl = {https://ui.adsabs.harvard.edu/abs/2017ApJ...850...40R}
}

@ARTICLE{Birzan2004,
       author = {{B{\^\i}rzan}, L. and {Rafferty}, D.~A. and {McNamara}, B.~R. and {Wise}, M.~W. and {Nulsen}, P.~E.~J.},
        title = "{A Systematic Study of Radio-induced X-Ray Cavities in Clusters, Groups, and Galaxies}",
      journal = {\apj},
         year = 2004,
        month = jun,
       volume = {607},
       number = {2},
        pages = {800-809},
          doi = {10.1086/383519},
archivePrefix = {arXiv},
       eprint = {astro-ph/0402348},
 primaryClass = {astro-ph},
       adsurl = {https://ui.adsabs.harvard.edu/abs/2004ApJ...607..800B}
}

@ARTICLE{Fabian2012,
       author = {{Fabian}, A.~C.},
        title = "{Observational Evidence of Active Galactic Nuclei Feedback}",
      journal = {\araa},
         year = 2012,
        month = sep,
       volume = {50},
        pages = {455-489},
          doi = {10.1146/annurev-astro-081811-125521},
archivePrefix = {arXiv},
       eprint = {1204.4114},
 primaryClass = {astro-ph.CO},
       adsurl = {https://ui.adsabs.harvard.edu/abs/2012ARA&A..50..455F}
}

@ARTICLE{Forster2020,
       author = {{F{\"o}rster Schreiber}, Natascha M. and {Wuyts}, Stijn},
        title = "{Star-Forming Galaxies at Cosmic Noon}",
      journal = {\araa},
         year = 2020,
        month = aug,
       volume = {58},
        pages = {661-725},
          doi = {10.1146/annurev-astro-032620-021910},
archivePrefix = {arXiv},
       eprint = {2010.10171},
 primaryClass = {astro-ph.GA},
       adsurl = {https://ui.adsabs.harvard.edu/abs/2020ARA&A..58..661F}
}

@ARTICLE{Feretti2012,
       author = {{Feretti}, Luigina and {Giovannini}, Gabriele and {Govoni}, Federica and {Murgia}, Matteo},
        title = "{Clusters of galaxies: observational properties of the diffuse radio emission}",
      journal = {\aapr},
         year = 2012,
        month = may,
       volume = {20},
          eid = {54},
        pages = {54},
          doi = {10.1007/s00159-012-0054-z},
archivePrefix = {arXiv},
       eprint = {1205.1919},
 primaryClass = {astro-ph.CO},
       adsurl = {https://ui.adsabs.harvard.edu/abs/2012A&ARv..20...54F}
}

@ARTICLE{Zinger2020,
       author = {{Zinger}, Elad and {Pillepich}, Annalisa and {Nelson}, Dylan and {Weinberger}, Rainer and {Pakmor}, R{\"u}diger and {Springel}, Volker and {Hernquist}, Lars and {Marinacci}, Federico and {Vogelsberger}, Mark},
        title = "{Ejective and preventative: the IllustrisTNG black hole feedback and its effects on the thermodynamics of the gas within and around galaxies}",
      journal = {MNRAS},
         year = 2020,
        month = nov,
       volume = {499},
       number = {1},
        pages = {768-792},
          doi = {10.1093/mnras/staa2607},
archivePrefix = {arXiv},
       eprint = {2004.06132},
 primaryClass = {astro-ph.GA},
       adsurl = {https://ui.adsabs.harvard.edu/abs/2020MNRAS.499..768Z}
}

@article{Davies2019,
    author = {Davies, Jonathan J and Crain, Robert A and Oppenheimer, Benjamin D and Schaye, Joop},
    title = {The quenching and morphological evolution of central galaxies is facilitated by the feedback-driven expulsion of circumgalactic gas},
    journal = {MNRAS},
    volume = {491},
    number = {3},
    pages = {4462-4480},
    year = {2019},
    month = {11},
    issn = {0035-8711},
    doi = {10.1093/mnras/stz3201},
    url = {https://doi.org/10.1093/mnras/stz3201},
    eprint = {https://academic.oup.com/mnras/article-pdf/491/3/4462/31559401/stz3201.pdf},
}

@article{Liu2012,
   title={Star formation activities in early-type brightest cluster galaxies: Star formation activities in early-type BCGs},
   volume={423},
   ISSN={0035-8711},
   url={http://dx.doi.org/10.1111/j.1365-2966.2012.20886.x},
   DOI={10.1111/j.1365-2966.2012.20886.x},
   number={1},
   journal={MNRAS},
   publisher={Oxford University Press (OUP)},
   author={Liu, F. S. and Mao, Shude and Meng, X. M.},
   year={2012},
   month=apr, pages={422–436} }

@ARTICLE{kurinchi-vendhan2024,
       author = {{Kurinchi-Vendhan}, Shalini and {Farcy}, Marion and {Hirschmann}, Michaela and {Valentino}, Francesco},
        title = "{On the origin of star formation quenching in massive galaxies at z {\ensuremath{\gtrsim}} 3 in the cosmological simulations IllustrisTNG}",
      journal = {\mnras},
         year = 2024,
        month = nov,
       volume = {534},
       number = {4},
        pages = {3974-3988},
          doi = {10.1093/mnras/stae2297},
archivePrefix = {arXiv},
       eprint = {2310.03083},
 primaryClass = {astro-ph.GA},
       adsurl = {https://ui.adsabs.harvard.edu/abs/2024MNRAS.534.3974K}
}

@ARTICLE{truong2021,
       author = {{Truong}, Nhut and {Pillepich}, Annalisa and {Werner}, Norbert},
        title = "{Correlations between supermassive black holes and hot gas atmospheres in IllustrisTNG and X-ray observations}",
      journal = {\mnras},
         year = 2021,
        month = feb,
       volume = {501},
       number = {2},
        pages = {2210-2230},
          doi = {10.1093/mnras/staa3880},
archivePrefix = {arXiv},
       eprint = {2009.06634},
 primaryClass = {astro-ph.CO},
       adsurl = {https://ui.adsabs.harvard.edu/abs/2021MNRAS.501.2210T}
}

@ARTICLE{Varela2009,
       author = {{Varela}, J. and {D'Onofrio}, M. and {Marmo}, C. and {Fasano}, G. and {Bettoni}, D. and {Cava}, A. and {Couch}, W.~J. and {Dressler}, A. and {Kj{\ae}rgaard}, P. and {Moles}, M. and {Pignatelli}, E. and {Poggianti}, B.~M. and {Valentinuzzi}, T.},
        title = "{WINGS: A WIde-field Nearby Galaxy-cluster Survey. II. Deep optical photometry of 77 nearby clusters}",
      journal = {\aap},
         year = 2009,
        month = apr,
       volume = {497},
       number = {3},
        pages = {667-676},
          doi = {10.1051/0004-6361/200809876},
archivePrefix = {arXiv},
       eprint = {0902.0612},
 primaryClass = {astro-ph.CO},
       adsurl = {https://ui.adsabs.harvard.edu/abs/2009A&A...497..667V}
}

@ARTICLE{Lehle2025,
       author = {{Lehle}, Katrin and {Nelson}, Dylan and {Pillepich}, Annalisa},
        title = "{What drives cluster cool-core transformations? A population-level analysis using TNG-Cluster}",
      journal = {\aap},
         year = 2025,
        month = oct,
       volume = {702},
          eid = {A25},
        pages = {A25},
          doi = {10.1051/0004-6361/202554374},
archivePrefix = {arXiv},
       eprint = {2503.01969},
 primaryClass = {astro-ph.GA},
       adsurl = {https://ui.adsabs.harvard.edu/abs/2025A&A...702A..25L}
}

@ARTICLE{Mihos1996,
       author = {{Mihos}, J. Christopher and {Hernquist}, Lars},
        title = "{Gasdynamics and Starbursts in Major Mergers}",
      journal = {\apj},
         year = 1996,
        month = jun,
       volume = {464},
        pages = {641},
          doi = {10.1086/177353},
archivePrefix = {arXiv},
       eprint = {astro-ph/9512099},
 primaryClass = {astro-ph},
       adsurl = {https://ui.adsabs.harvard.edu/abs/1996ApJ...464..641M}
}

@ARTICLE{di_matteo2005,
       author = {{Di Matteo}, Tiziana and {Springel}, Volker and {Hernquist}, Lars},
        title = "{Energy input from quasars regulates the growth and activity of black holes and their host galaxies}",
      journal = {\nat},
         year = 2005,
        month = feb,
       volume = {433},
       number = {7026},
        pages = {604-607},
          doi = {10.1038/nature03335},
archivePrefix = {arXiv},
       eprint = {astro-ph/0502199},
 primaryClass = {astro-ph},
       adsurl = {https://ui.adsabs.harvard.edu/abs/2005Natur.433..604D}
}

@ARTICLE{Pawlik2018,
       author = {{Pawlik}, M.~M. and {Taj Aldeen}, L. and {Wild}, V. and {Mendez-Abreu}, J. and {Lah{\'e}n}, N. and {Johansson}, P.~H. and {Jimenez}, N. and {Lucas}, W. and {Zheng}, Y. and {Walcher}, C.~J. and {Rowlands}, K.},
        title = "{The origins of post-starburst galaxies at z < 0.05}",
      journal = {\mnras},
         year = 2018,
        month = jun,
       volume = {477},
       number = {2},
        pages = {1708-1743},
          doi = {10.1093/mnras/sty589},
archivePrefix = {arXiv},
       eprint = {1803.02201},
 primaryClass = {astro-ph.GA},
       adsurl = {https://ui.adsabs.harvard.edu/abs/2018MNRAS.477.1708P}
}

@ARTICLE{hernquist1989,
       author = {{Hernquist}, Lars},
        title = "{Tidal triggering of starbursts and nuclear activity in galaxies}",
      journal = {\nat},
         year = 1989,
        month = aug,
       volume = {340},
       number = {6236},
        pages = {687-691},
          doi = {10.1038/340687a0},
       adsurl = {https://ui.adsabs.harvard.edu/abs/1989Natur.340..687H}
}

@ARTICLE{webb2015,
       author = {{Webb}, Tracy M.~A. and {Muzzin}, Adam and {Noble}, Allison and {Bonaventura}, Nina and {Geach}, James and {Hezaveh}, Yashar and {Lidman}, Chris and {Wilson}, Gillian and {Yee}, H.~K.~C. and {Surace}, Jason and {Shupe}, David},
        title = "{The Star Formation History of BCGs to z = 1.8 from the SpARCS/SWIRE Survey: Evidence for Significant In Situ Star Formation at High Redshift}",
      journal = {\apj},
         year = 2015,
        month = dec,
       volume = {814},
       number = {2},
          eid = {96},
        pages = {96},
          doi = {10.1088/0004-637X/814/2/96},
archivePrefix = {arXiv},
       eprint = {1508.07302},
 primaryClass = {astro-ph.GA},
       adsurl = {https://ui.adsabs.harvard.edu/abs/2015ApJ...814...96W}
}

@ARTICLE{Belli2019,
       author = {{Belli}, Sirio and {Newman}, Andrew B. and {Ellis}, Richard S.},
        title = "{MOSFIRE Spectroscopy of Quiescent Galaxies at 1.5 < z < 2.5. II. Star Formation Histories and Galaxy Quenching}",
      journal = {\apj},
         year = 2019,
        month = mar,
       volume = {874},
       number = {1},
          eid = {17},
        pages = {17},
          doi = {10.3847/1538-4357/ab07af},
archivePrefix = {arXiv},
       eprint = {1810.00008},
 primaryClass = {astro-ph.GA},
       adsurl = {https://ui.adsabs.harvard.edu/abs/2019ApJ...874...17B}
}

@ARTICLE{Carnall2019,
       author = {{Carnall}, A.~C. and {McLure}, R.~J. and {Dunlop}, J.~S. and {Cullen}, F. and {McLeod}, D.~J. and {Wild}, V. and {Johnson}, B.~D. and {Appleby}, S. and {Dav{\'e}}, R. and {Amorin}, R. and {Bolzonella}, M. and {Castellano}, M. and {Cimatti}, A. and {Cucciati}, O. and {Gargiulo}, A. and {Garilli}, B. and {Marchi}, F. and {Pentericci}, L. and {Pozzetti}, L. and {Schreiber}, C. and {Talia}, M. and {Zamorani}, G.},
        title = "{The VANDELS survey: the star-formation histories of massive quiescent galaxies at 1.0 < z < 1.3}",
      journal = {\mnras},
         year = 2019,
        month = nov,
       volume = {490},
       number = {1},
        pages = {417-439},
          doi = {10.1093/mnras/stz2544},
archivePrefix = {arXiv},
       eprint = {1903.11082},
 primaryClass = {astro-ph.GA},
       adsurl = {https://ui.adsabs.harvard.edu/abs/2019MNRAS.490..417C}
}

@ARTICLE{Weaver2023,
       author = {{Weaver}, J.~R. and {Davidzon}, I. and {Toft}, S. and {Ilbert}, O. and {McCracken}, H.~J. and {Gould}, K.~M.~L. and {Jespersen}, C.~K. and {Steinhardt}, C. and {Lagos}, C.~D.~P. and {Capak}, P.~L. and {Casey}, C.~M. and {Chartab}, N. and {Faisst}, A.~L. and {Hayward}, C.~C. and {Kartaltepe}, J.~S. and {Kauffmann}, O.~B. and {Koekemoer}, A.~M. and {Kokorev}, V. and {Laigle}, C. and {Liu}, D. and {Long}, A. and {Magdis}, G.~E. and {McPartland}, C.~J.~R. and {Milvang-Jensen}, B. and {Mobasher}, B. and {Moneti}, A. and {Peng}, Y. and {Sanders}, D.~B. and {Shuntov}, M. and {Sneppen}, A. and {Valentino}, F. and {Zalesky}, L. and {Zamorani}, G.},
        title = "{COSMOS2020: The galaxy stellar mass function. The assembly and star formation cessation of galaxies at 0.2< z {\ensuremath{\leq}} 7.5}",
      journal = {\aap},
         year = 2023,
        month = sep,
       volume = {677},
          eid = {A184},
        pages = {A184},
          doi = {10.1051/0004-6361/202245581},
archivePrefix = {arXiv},
       eprint = {2212.02512},
 primaryClass = {astro-ph.GA},
       adsurl = {https://ui.adsabs.harvard.edu/abs/2023A&A...677A.184W}
}

@ARTICLE{McDonald2016,
       author = {{McDonald}, M. and {Stalder}, B. and {Bayliss}, M. and {Allen}, S.~W. and {Applegate}, D.~E. and {Ashby}, M.~L.~N. and {Bautz}, M. and {Benson}, B.~A. and {Bleem}, L.~E. and {Brodwin}, M. and {Carlstrom}, J.~E. and {Chiu}, I. and {Desai}, S. and {Gonzalez}, A.~H. and {Hlavacek-Larrondo}, J. and {Holzapfel}, W.~L. and {Marrone}, D.~P. and {Miller}, E.~D. and {Reichardt}, C.~L. and {Saliwanchik}, B.~R. and {Saro}, A. and {Schrabback}, T. and {Stanford}, S.~A. and {Stark}, A.~A. and {Vieira}, J.~D. and {Zenteno}, A.},
        title = "{Star-forming Brightest Cluster Galaxies at 0.25 < z < 1.25: A Transitioning Fuel Supply}",
      journal = {\apj},
         year = 2016,
        month = feb,
       volume = {817},
       number = {2},
          eid = {86},
        pages = {86},
          doi = {10.3847/0004-637X/817/2/86},
archivePrefix = {arXiv},
       eprint = {1508.06283},
 primaryClass = {astro-ph.GA},
       adsurl = {https://ui.adsabs.harvard.edu/abs/2016ApJ...817...86M}
}

@ARTICLE{Slob2024,
       author = {{Slob}, Martje and {Kriek}, Mariska and {Beverage}, Aliza G. and {Suess}, Katherine A. and {Barro}, Guillermo and {Bezanson}, Rachel and {Brammer}, Gabriel and {Cheng}, Chloe M. and {Conroy}, Charlie and {de Graaff}, Anna and {F{\"o}rster Schreiber}, Natascha M. and {Franx}, Marijn and {Lorenz}, Brian and {Mancera Pi{\~n}a}, Pavel E. and {Marchesini}, Danilo and {Muzzin}, Adam and {Newman}, Andrew B. and {Price}, Sedona H. and {Shapley}, Alice E. and {Stefanon}, Mauro and {van Dokkum}, Pieter and {Weisz}, Daniel R.},
        title = "{The JWST-SUSPENSE Ultradeep Spectroscopic Program: Survey Overview and Star Formation Histories of Quiescent Galaxies at 1 < z < 3}",
      journal = {\apj},
         year = 2024,
        month = oct,
       volume = {973},
       number = {2},
          eid = {131},
        pages = {131},
          doi = {10.3847/1538-4357/ad65ff},
archivePrefix = {arXiv},
       eprint = {2404.12432},
 primaryClass = {astro-ph.GA},
       adsurl = {https://ui.adsabs.harvard.edu/abs/2024ApJ...973..131S}
}

@ARTICLE{Belli2017,
       author = {{Belli}, Sirio and {Genzel}, Reinhard and {F{\"o}rster Schreiber}, Natascha M. and {Wisnioski}, Emily and {Wilman}, David J. and {Wuyts}, Stijn and {Mendel}, J. Trevor and {Beifiori}, Alessandra and {Bender}, Ralf and {Brammer}, Gabriel B. and {Burkert}, Andreas and {Chan}, Jeffrey and {Davies}, Rebecca L. and {Davies}, Ric and {Fabricius}, Maximilian and {Fossati}, Matteo and {Galametz}, Audrey and {Lang}, Philipp and {Lutz}, Dieter and {Momcheva}, Ivelina G. and {Nelson}, Erica J. and {Saglia}, Roberto P. and {Tacconi}, Linda J. and {Tadaki}, Ken-ichi and {{\"U}bler}, Hannah and {van Dokkum}, Pieter},
        title = "{KMOS$^{3D}$ Reveals Low-level Star Formation Activity in Massive Quiescent Galaxies at 0.7 < z < 2.7}",
      journal = {\apjl},
         year = 2017,
        month = may,
       volume = {841},
       number = {1},
          eid = {L6},
        pages = {L6},
          doi = {10.3847/2041-8213/aa70e5},
archivePrefix = {arXiv},
       eprint = {1703.07778},
 primaryClass = {astro-ph.GA},
       adsurl = {https://ui.adsabs.harvard.edu/abs/2017ApJ...841L...6B}
}

@ARTICLE{Peng2010,
       author = {{Peng}, Ying-jie and {Lilly}, Simon J. and {Kova{\v{c}}}, Katarina and {Bolzonella}, Micol and {Pozzetti}, Lucia and {Renzini}, Alvio and {Zamorani}, Gianni and {Ilbert}, Olivier and {Knobel}, Christian and {Iovino}, Angela and {Maier}, Christian and {Cucciati}, Olga and {Tasca}, Lidia and {Carollo}, C. Marcella and {Silverman}, John and {Kampczyk}, Pawel and {de Ravel}, Loic and {Sanders}, David and {Scoville}, Nicholas and {Contini}, Thierry and {Mainieri}, Vincenzo and {Scodeggio}, Marco and {Kneib}, Jean-Paul and {Le F{\`e}vre}, Olivier and {Bardelli}, Sandro and {Bongiorno}, Angela and {Caputi}, Karina and {Coppa}, Graziano and {de la Torre}, Sylvain and {Franzetti}, Paolo and {Garilli}, Bianca and {Lamareille}, Fabrice and {Le Borgne}, Jean-Francois and {Le Brun}, Vincent and {Mignoli}, Marco and {Perez Montero}, Enrique and {Pello}, Roser and {Ricciardelli}, Elena and {Tanaka}, Masayuki and {Tresse}, Laurence and {Vergani}, Daniela and {Welikala}, Niraj and {Zucca}, Elena and {Oesch}, Pascal and {Abbas}, Ummi and {Barnes}, Luke and {Bordoloi}, Rongmon and {Bottini}, Dario and {Cappi}, Alberto and {Cassata}, Paolo and {Cimatti}, Andrea and {Fumana}, Marco and {Hasinger}, Gunther and {Koekemoer}, Anton and {Leauthaud}, Alexei and {Maccagni}, Dario and {Marinoni}, Christian and {McCracken}, Henry and {Memeo}, Pierdomenico and {Meneux}, Baptiste and {Nair}, Preethi and {Porciani}, Cristiano and {Presotto}, Valentina and {Scaramella}, Roberto},
        title = "{Mass and Environment as Drivers of Galaxy Evolution in SDSS and zCOSMOS and the Origin of the Schechter Function}",
      journal = {\apj},
         year = 2010,
        month = sep,
       volume = {721},
       number = {1},
        pages = {193-221},
          doi = {10.1088/0004-637X/721/1/193},
archivePrefix = {arXiv},
       eprint = {1003.4747},
 primaryClass = {astro-ph.CO},
       adsurl = {https://ui.adsabs.harvard.edu/abs/2010ApJ...721..193P}
}

@ARTICLE{Montenegro-Taborda2023,
       author = {{Montenegro-Taborda}, Daniel and {Rodriguez-Gomez}, Vicente and {Pillepich}, Annalisa and {Avila-Reese}, Vladimir and {Sales}, Laura V. and {Rodr{\'\i}guez-Puebla}, Aldo and {Hernquist}, Lars},
        title = "{The growth of brightest cluster galaxies in the TNG300 simulation: dissecting the contributions from mergers and in situ star formation}",
      journal = {\mnras},
         year = 2023,
        month = may,
       volume = {521},
       number = {1},
        pages = {800-817},
          doi = {10.1093/mnras/stad586},
archivePrefix = {arXiv},
       eprint = {2302.10943},
 primaryClass = {astro-ph.GA},
       adsurl = {https://ui.adsabs.harvard.edu/abs/2023MNRAS.521..800M}
}

@ARTICLE{Aragon-Salamanca1998,
       author = {{Aragon-Salamanca}, Alfonso and {Baugh}, Carlton M. and {Kauffmann}, Guinevere},
        title = "{The K-band Hubble diagram for the brightest cluster galaxies: a test of hierarchical galaxy formation models}",
      journal = {\mnras},
         year = 1998,
        month = jun,
       volume = {297},
       number = {2},
        pages = {427-434},
          doi = {10.1046/j.1365-8711.1998.01495.x},
archivePrefix = {arXiv},
       eprint = {astro-ph/9801277},
 primaryClass = {astro-ph},
       adsurl = {https://ui.adsabs.harvard.edu/abs/1998MNRAS.297..427A}
}

@ARTICLE{Pipino2009,
       author = {{Pipino}, A. and {Kaviraj}, S. and {Bildfell}, C. and {Babul}, A. and {Hoekstra}, H. and {Silk}, J.},
        title = "{Evidence for recent star formation in BCGs: a correspondence between blue cores and UV excess}",
      journal = {\mnras},
         year = 2009,
        month = may,
       volume = {395},
       number = {1},
        pages = {462-471},
          doi = {10.1111/j.1365-2966.2009.14534.x},
archivePrefix = {arXiv},
       eprint = {0807.2760},
 primaryClass = {astro-ph},
       adsurl = {https://ui.adsabs.harvard.edu/abs/2009MNRAS.395..462P}
}

@ARTICLE{Wang2010,
       author = {{Wang}, Jing and {Overzier}, Roderik and {Kauffmann}, Guinevere and {von der Linden}, Anja and {Kong}, Xu},
        title = "{The UV-optical colours of brightest cluster galaxies in optically and X-ray selected clusters}",
      journal = {\mnras},
         year = 2010,
        month = jan,
       volume = {401},
       number = {1},
        pages = {433-444},
          doi = {10.1111/j.1365-2966.2009.15653.x},
archivePrefix = {arXiv},
       eprint = {0909.1196},
 primaryClass = {astro-ph.GA},
       adsurl = {https://ui.adsabs.harvard.edu/abs/2010MNRAS.401..433W}
}

@ARTICLE{Somerville2015,
       author = {{Somerville}, Rachel S. and {Dav{\'e}}, Romeel},
        title = "{Physical Models of Galaxy Formation in a Cosmological Framework}",
      journal = {\araa},
         year = 2015,
        month = aug,
       volume = {53},
        pages = {51-113},
          doi = {10.1146/annurev-astro-082812-140951},
archivePrefix = {arXiv},
       eprint = {1412.2712},
 primaryClass = {astro-ph.GA},
       adsurl = {https://ui.adsabs.harvard.edu/abs/2015ARA&A..53...51S}
}

@ARTICLE{Dubois2016,
       author = {{Dubois}, Yohan and {Peirani}, S{\'e}bastien and {Pichon}, Christophe and {Devriendt}, Julien and {Gavazzi}, Rapha{\"e}l and {Welker}, Charlotte and {Volonteri}, Marta},
        title = "{The HORIZON-AGN simulation: morphological diversity of galaxies promoted by AGN feedback}",
      journal = {\mnras},
         year = 2016,
        month = dec,
       volume = {463},
       number = {4},
        pages = {3948-3964},
          doi = {10.1093/mnras/stw2265},
archivePrefix = {arXiv},
       eprint = {1606.03086},
 primaryClass = {astro-ph.GA},
       adsurl = {https://ui.adsabs.harvard.edu/abs/2016MNRAS.463.3948D}
}

@ARTICLE{Ni2025,
       author = {{Ni}, Yueying and {Chen}, Nianyi and {Zhou}, Yihao and {Park}, Minjung and {Yang}, Yanhui and {Di Matteo}, Tiziana and {Bird}, Simeon and {Croft}, Rupert},
        title = "{The Astrid Simulation: Evolution of Black Holes and Galaxies to z = 0.5 and Different Evolution Pathways for Galaxy Quenching}",
      journal = {\apj},
         year = 2025,
        month = sep,
       volume = {990},
       number = {2},
          eid = {120},
        pages = {120},
          doi = {10.3847/1538-4357/adf3a7},
archivePrefix = {arXiv},
       eprint = {2409.10666},
 primaryClass = {astro-ph.GA},
       adsurl = {https://ui.adsabs.harvard.edu/abs/2025ApJ...990..120N}
}

@ARTICLE{Schaye2015,
       author = {{Schaye}, Joop and {Crain}, Robert A. and {Bower}, Richard G. and {Furlong}, Michelle and {Schaller}, Matthieu and {Theuns}, Tom and {Dalla Vecchia}, Claudio and {Frenk}, Carlos S. and {McCarthy}, I.~G. and {Helly}, John C. and {Jenkins}, Adrian and {Rosas-Guevara}, Y.~M. and {White}, Simon D.~M. and {Baes}, Maarten and {Booth}, C.~M. and {Camps}, Peter and {Navarro}, Julio F. and {Qu}, Yan and {Rahmati}, Alireza and {Sawala}, Till and {Thomas}, Peter A. and {Trayford}, James},
        title = "{The EAGLE project: simulating the evolution and assembly of galaxies and their environments}",
      journal = {\mnras},
         year = 2015,
        month = jan,
       volume = {446},
       number = {1},
        pages = {521-554},
          doi = {10.1093/mnras/stu2058},
archivePrefix = {arXiv},
       eprint = {1407.7040},
 primaryClass = {astro-ph.GA},
       adsurl = {https://ui.adsabs.harvard.edu/abs/2015MNRAS.446..521S}
}

@ARTICLE{Crain2015,
       author = {{Crain}, Robert A. and {Schaye}, Joop and {Bower}, Richard G. and {Furlong}, Michelle and {Schaller}, Matthieu and {Theuns}, Tom and {Dalla Vecchia}, Claudio and {Frenk}, Carlos S. and {McCarthy}, Ian G. and {Helly}, John C. and {Jenkins}, Adrian and {Rosas-Guevara}, Yetli M. and {White}, Simon D.~M. and {Trayford}, James W.},
        title = "{The EAGLE simulations of galaxy formation: calibration of subgrid physics and model variations}",
      journal = {\mnras},
         year = 2015,
        month = jun,
       volume = {450},
       number = {2},
        pages = {1937-1961},
          doi = {10.1093/mnras/stv725},
archivePrefix = {arXiv},
       eprint = {1501.01311},
 primaryClass = {astro-ph.GA},
       adsurl = {https://ui.adsabs.harvard.edu/abs/2015MNRAS.450.1937C}
}

@ARTICLE{Scharre2024,
       author = {{Scharr{\'e}}, Lucie and {Sorini}, Daniele and {Dav{\'e}}, Romeel},
        title = "{The effects of stellar and AGN feedback on the cosmic star formation history in the SIMBA simulations}",
      journal = {\mnras},
         year = 2024,
        month = oct,
       volume = {534},
       number = {1},
        pages = {361-383},
          doi = {10.1093/mnras/stae2098},
archivePrefix = {arXiv},
       eprint = {2404.07252},
 primaryClass = {astro-ph.GA},
       adsurl = {https://ui.adsabs.harvard.edu/abs/2024MNRAS.534..361S}
}

@ARTICLE{Nelson2019_tng50,
       author = {{Nelson}, Dylan and {Pillepich}, Annalisa and {Springel}, Volker and {Pakmor}, R{\"u}diger and {Weinberger}, Rainer and {Genel}, Shy and {Torrey}, Paul and {Vogelsberger}, Mark and {Marinacci}, Federico and {Hernquist}, Lars},
        title = "{First results from the TNG50 simulation: galactic outflows driven by supernovae and black hole feedback}",
      journal = {\mnras},
         year = 2019,
        month = dec,
       volume = {490},
       number = {3},
        pages = {3234-3261},
          doi = {10.1093/mnras/stz2306},
archivePrefix = {arXiv},
       eprint = {1902.05554},
 primaryClass = {astro-ph.GA},
       adsurl = {https://ui.adsabs.harvard.edu/abs/2019MNRAS.490.3234N}
}

@ARTICLE{Pillepich2019_tng50,
       author = {{Pillepich}, Annalisa and {Nelson}, Dylan and {Springel}, Volker and {Pakmor}, R{\"u}diger and {Torrey}, Paul and {Weinberger}, Rainer and {Vogelsberger}, Mark and {Marinacci}, Federico and {Genel}, Shy and {van der Wel}, Arjen and {Hernquist}, Lars},
        title = "{First results from the TNG50 simulation: the evolution of stellar and gaseous discs across cosmic time}",
      journal = {\mnras},
         year = 2019,
        month = dec,
       volume = {490},
       number = {3},
        pages = {3196-3233},
          doi = {10.1093/mnras/stz2338},
archivePrefix = {arXiv},
       eprint = {1902.05553},
 primaryClass = {astro-ph.GA},
       adsurl = {https://ui.adsabs.harvard.edu/abs/2019MNRAS.490.3196P}
}

@ARTICLE{Donnari2021obs,
       author = {{Donnari}, Martina and {Pillepich}, Annalisa and {Nelson}, Dylan and {Marinacci}, Federico and {Vogelsberger}, Mark and {Hernquist}, Lars},
        title = "{Quenched fractions in the IllustrisTNG simulations: comparison with observations and other theoretical models}",
      journal = {\mnras},
         year = 2021,
        month = oct,
       volume = {506},
       number = {4},
        pages = {4760-4780},
          doi = {10.1093/mnras/stab1950},
archivePrefix = {arXiv},
       eprint = {2008.00004},
 primaryClass = {astro-ph.GA},
       adsurl = {https://ui.adsabs.harvard.edu/abs/2021MNRAS.506.4760D}
}

@ARTICLE{Davies2020,
       author = {{Davies}, Jonathan J. and {Crain}, Robert A. and {Oppenheimer}, Benjamin D. and {Schaye}, Joop},
        title = "{The quenching and morphological evolution of central galaxies is facilitated by the feedback-driven expulsion of circumgalactic gas}",
      journal = {\mnras},
         year = 2020,
        month = jan,
       volume = {491},
       number = {3},
        pages = {4462-4480},
          doi = {10.1093/mnras/stz3201},
archivePrefix = {arXiv},
       eprint = {1908.11380},
 primaryClass = {astro-ph.GA},
       adsurl = {https://ui.adsabs.harvard.edu/abs/2020MNRAS.491.4462D}
}

@ARTICLE{EricaNelson2021,
       author = {{Nelson}, Erica J. and {Tacchella}, Sandro and {Diemer}, Benedikt and {Leja}, Joel and {Hernquist}, Lars and {Whitaker}, Katherine E. and {Weinberger}, Rainer and {Pillepich}, Annalisa and {Nelson}, Dylan and {Terrazas}, Bryan A. and {Nevin}, Rebecca and {Brammer}, Gabriel B. and {Burkhart}, Blakesley and {Cochrane}, Rachel K. and {van Dokkum}, Pieter and {Johnson}, Benjamin D. and {Marinacci}, Federico and {Mowla}, Lamiya and {Pakmor}, R{\"u}diger and {Skelton}, Rosalind E. and {Speagle}, Joshua and {Springel}, Volker and {Torrey}, Paul and {Vogelsberger}, Mark and {Wuyts}, Stijn},
        title = "{Spatially resolved star formation and inside-out quenching in the TNG50 simulation and 3D-HST observations}",
      journal = {\mnras},
         year = 2021,
        month = nov,
       volume = {508},
       number = {1},
        pages = {219-235},
          doi = {10.1093/mnras/stab2131},
archivePrefix = {arXiv},
       eprint = {2101.12212},
 primaryClass = {astro-ph.GA},
       adsurl = {https://ui.adsabs.harvard.edu/abs/2021MNRAS.508..219N}
}

@ARTICLE{Vogelsberger2014,
       author = {{Vogelsberger}, Mark and {Genel}, Shy and {Springel}, Volker and {Torrey}, Paul and {Sijacki}, Debora and {Xu}, Dandan and {Snyder}, Greg and {Nelson}, Dylan and {Hernquist}, Lars},
        title = "{Introducing the Illustris Project: simulating the coevolution of dark and visible matter in the Universe}",
      journal = {\mnras},
         year = 2014,
        month = oct,
       volume = {444},
       number = {2},
        pages = {1518-1547},
          doi = {10.1093/mnras/stu1536},
archivePrefix = {arXiv},
       eprint = {1405.2921},
 primaryClass = {astro-ph.CO},
       adsurl = {https://ui.adsabs.harvard.edu/abs/2014MNRAS.444.1518V}
}

@ARTICLE{Pillepich2021,
       author = {{Pillepich}, Annalisa and {Nelson}, Dylan and {Truong}, Nhut and {Weinberger}, Rainer and {Martin-Navarro}, Ignacio and {Springel}, Volker and {Faber}, Sandy M. and {Hernquist}, Lars},
        title = "{X-ray bubbles in the circumgalactic medium of TNG50 Milky Way- and M31-like galaxies: signposts of supermassive black hole activity}",
      journal = {\mnras},
         year = 2021,
        month = dec,
       volume = {508},
       number = {4},
        pages = {4667-4695},
          doi = {10.1093/mnras/stab2779},
archivePrefix = {arXiv},
       eprint = {2105.08062},
 primaryClass = {astro-ph.GA},
       adsurl = {https://ui.adsabs.harvard.edu/abs/2021MNRAS.508.4667P}
}

@ARTICLE{Cerulo2019,
       author = {{Cerulo}, P. and {Orellana}, G.~A. and {Covone}, G.},
        title = "{The evolution of brightest cluster galaxies in the nearby Universe - I. Colours and stellar masses from the Sloan Digital Sky Survey and Wide Infrared Survey Explorer}",
      journal = {\mnras},
         year = 2019,
        month = aug,
       volume = {487},
       number = {3},
        pages = {3759-3775},
          doi = {10.1093/mnras/stz1495},
archivePrefix = {arXiv},
       eprint = {1905.12117},
 primaryClass = {astro-ph.GA},
       adsurl = {https://ui.adsabs.harvard.edu/abs/2019MNRAS.487.3759C}
}

@ARTICLE{Lin2013,
       author = {{Lin}, Yen-Ting and {Brodwin}, Mark and {Gonzalez}, Anthony H. and {Bode}, Paul and {Eisenhardt}, Peter R.~M. and {Stanford}, S.~A. and {Vikhlinin}, Alexey},
        title = "{The Stellar Mass Growth of Brightest Cluster Galaxies in the IRAC Shallow Cluster Survey}",
      journal = {\apj},
         year = 2013,
        month = jul,
       volume = {771},
       number = {1},
          eid = {61},
        pages = {61},
          doi = {10.1088/0004-637X/771/1/61},
archivePrefix = {arXiv},
       eprint = {1305.5254},
 primaryClass = {astro-ph.CO},
       adsurl = {https://ui.adsabs.harvard.edu/abs/2013ApJ...771...61L}
}

@ARTICLE{Coziol2009,
       author = {{Coziol}, R. and {Andernach}, H. and {Caretta}, C.~A. and {Alamo-Mart{\'\i}nez}, K.~A. and {Tago}, E.},
        title = "{The Dynamical State of Brightest Cluster Galaxies and The Formation of Clusters}",
      journal = {\aj},
         year = 2009,
        month = jun,
       volume = {137},
       number = {6},
        pages = {4795-4809},
          doi = {10.1088/0004-6256/137/6/4795},
archivePrefix = {arXiv},
       eprint = {0903.2360},
 primaryClass = {astro-ph.CO},
       adsurl = {https://ui.adsabs.harvard.edu/abs/2009AJ....137.4795C}
}

@ARTICLE{Lauer2014,
       author = {{Lauer}, Tod R. and {Postman}, Marc and {Strauss}, Michael A. and {Graves}, Genevieve J. and {Chisari}, Nora E.},
        title = "{Brightest Cluster Galaxies at the Present Epoch}",
      journal = {\apj},
         year = 2014,
        month = dec,
       volume = {797},
       number = {2},
          eid = {82},
        pages = {82},
          doi = {10.1088/0004-637X/797/2/82},
archivePrefix = {arXiv},
       eprint = {1407.2260},
 primaryClass = {astro-ph.GA},
       adsurl = {https://ui.adsabs.harvard.edu/abs/2014ApJ...797...82L}
}

@ARTICLE{DePropris2021,
       author = {{De Propris}, Roberto and {West}, Michael J. and {Andrade-Santos}, Felipe and {Ragone-Figueroa}, Cinthia and {Rasia}, Elena and {Forman}, William and {Jones}, Christine and {Kipper}, Rain and {Borgani}, Stefano and {Lambas}, Diego Garc{\'\i}a and {Romashkova}, Elena A. and {Patra}, Kishore C.},
        title = "{Brightest cluster galaxies: the centre can(not?) hold}",
      journal = {\mnras},
         year = 2021,
        month = jan,
       volume = {500},
       number = {1},
        pages = {310-318},
          doi = {10.1093/mnras/staa3286},
archivePrefix = {arXiv},
       eprint = {2010.09617},
 primaryClass = {astro-ph.GA},
       adsurl = {https://ui.adsabs.harvard.edu/abs/2021MNRAS.500..310D}
}


\bsp	
\label{lastpage}
\end{document}